\documentclass{emulateapj}
\usepackage{color}
\begin{document}

\title{Untangling the Nature of Spatial Variations of Cold Dust Properties in Star Forming Galaxies}
\author{Allison Kirkpatrick\altaffilmark{1}, Daniela Calzetti\altaffilmark{1}, Robert Kennicutt\altaffilmark{2}, Maud Galametz\altaffilmark{3}, Karl Gordon\altaffilmark{4}, Brent Groves\altaffilmark{5}, Leslie Hunt\altaffilmark{6}, Daniel Dale\altaffilmark{7}, Joannah Hinz\altaffilmark{8}, Fatemeh Tabatabaei\altaffilmark{5}}
\altaffiltext{1}{Department of Astronomy, University of Massachusetts, Amherst, MA 01002, USA, kirkpatr@astro.umass.edu}
\altaffiltext{2}{Institute of Astronomy, University of Cambridge, Madingley Road, Cambridge, CB3 0HA, UK}
\altaffiltext{3}{European Southern Observatory, Karl-Schwarzchild-Str. 2, D-85748 Garching-bei-M\"{u}nchen, Germany}
\altaffiltext{4}{Space Telescope Science Institute, 3700 San Martin Drive, Baltimore, MD 21218, USA}
\altaffiltext{5}{Max-Planck Institut f\"{u}r Astronomie, K\"{o}nigstuhl 17, D-69117, Heidelberg, Germany}
\altaffiltext{6}{INAF-Osservatorio Astrofisico di Arcetri, Largo E. Fermi 5, I-50125 Firenze, Italy}
\altaffiltext{7}{Department of Physics \& Astronomy, University of Wyoming, Laramie, WY 82071, USA}
\altaffiltext{8}{MMT Observatory, University of Arizona, 933 N. Cherry Ave, Tucson, AZ 85721, USA}

\begin{abstract}
We investigate the far-infrared (IR) dust emission for 20 local star forming galaxies from the Key Insights on Nearby Galaxies: A Far-IR Survey with {\it Herschel} (KINGFISH) sample.
We model the far-IR/submillimeter spectral energy distribution (SED) using images from {\it Spitzer Space Telescope} and {\it Herschel Space Observatory}. We calculate the cold dust temperature ($T_c$) and emissivity ($\beta$) on a pixel by pixel basis (where each pixel ranges from $0.1-3$\,kpc$^2$) using a two temperature modified blackbody fitting routine. Our fitting method allows us to investigate the resolved
nature of temperature and emissivity variations by modeling from the galaxy centers to the outskirts (physical scales of $\sim15-50$ kpc, depending on the size of the galaxy). We fit each SED in two ways: (1) fit $T_c$ and $\beta$ simultaneously, (2) hold $\beta$ constant and fit $T_c$.
We compare $T_c$ and $\beta$ with star formation rates (calculated from $L_{{\rm H}\alpha}$ and $L_{24\,\mu{\rm m}}$),
the luminosity of the old stellar population (traced through $L_{3.6\,\mu{\rm m}}$), and the dust mass surface density (traced by 500\,$\mu$m luminosity, $L_{500}$). We find a significant trend between SFR/$L_{500}$ and $T_c$, implying that the flux of hard UV photons relative to the amount of dust is significantly
contributing to the heating of the cold, or diffuse, dust component. We also see a trend between $L_{3.6}/L_{500}$ and $\beta$, indicating that the old stellar population contributes to the heating
at far-IR/submillimeter wavelengths. Finally, we find that when $\beta$ is held constant, $T_c$ exhibits a strongly decreasing radial trend, illustrating that the shape of the far-IR
SED is changing radially through a galaxy, thus confirming on a sample almost double in size the trends observed in \citet{galametz2012}.
\end{abstract}

\section{Introduction}
Quantifying the properties of the interstellar medium (ISM) in a galaxy helps untangle the star formation history of that galaxy. The ISM is comprised of gas and dust particles.
The dust in the ISM is the product of several evolutionary processes. 
It is formed in the winds and ejecta of dying stars and supernova, and is further processed in the ISM before being incorporated into a new generation of stars.
The dust is heated by photons from young stars, the old stellar population,
and an active galactic nucleus (AGN), if present. Dust reradiates this absorbed light at infrared and submillimeter wavelengths. In recent decades, infrared space-based observatories have enabled 
detailed studies of dust emission that are not possible from the ground because of atmospheric absorption. 

If no AGN is present, newly formed stars are primarily responsible for heating the warm dust (T $\gtrsim 65\,$K), which radiates at wavelengths $\lambda\lesssim70\,\mu$m \citep{engelbracht2010,bendo2010}, while
cooler temperatures are associated with a dust component heated by the diffuse interstellar radiation field \citep[e.g.,][]{rowan1989, draine2003,tuffs2005, stevens2005}.
Photons produced in star forming sites heat the dust surrounding H{\sc ii} regions, but the presence of a widespread diffuse ionized medium in disk galaxies indicates that some of these
energetic photons must leak out of H{\sc ii} regions, so young stars could be indirectly responsible 
for heating the diffuse cold dust component as well \citep[e.g.,][]{reynolds1990a,reynolds1990b,wood2010,clemens2013,hughes2014}.

On the other hand, stellar bulges in spiral galaxies are observed to have warmer average dust temperatures than their disk components, a result which can be linked to heating by a dense old stellar population 
\citep{engelbracht2010,groves2012,draine2014}.
As for heating in the disk, several studies of local star forming galaxies have shown that temperature and far-IR colors of the cold dust decline as a function of radius \citep{bendo2010, galametz2012, hinz2012} which
could indicate that the old stellar population, also radially distributed, is responsible
for heating some of the cold dust \citep{boquien2011b}.
Conversely, the decline of dust temperatures radially
could simply reflect that star formation is also radially declining from the center of the galaxy outward \citep{galametz2012,hinz2012}. Indeed, \citet{law2011} use radiative transfer models to probe the far-IR emission of local, dusty galaxies and find that the old stellar population accounts for less than 20\% of the far-IR luminosity, and \citet{misi2001} apply dust heating models to edge-on spirals and determine that the old stellar population accounts for at most 40\% of the dust heating.

In addition to the dust temperatures and heating sources, the physical composition of the dust is seen to change from star forming regions to the diffuse ISM.
There is evidence that a substantial portion of grain growth occurs in the diffuse ISM \citep{draine2009}. A harsh interstellar radiation field can destroy the smallest grains,
polycyclic aromatic hydrocarbons (PAHs), which would result in these molecules being preferentially located in the photodissociation regions surrounding young stars \citep{helou2004}. Grain composition
could also vary with metallicity. For example, the Small Magellanic Cloud, a low metallicity galaxy, has a larger fraction of neutral and small PAHs than in the Milky Way
\citep{sandstrom2012}. On the other hand, the hard interstellar radiation field in low metallicity galaxies could be responsible for destroying the smallest PAH grains, resulting in a different grain size distribution than observed in the Milky Way \citep{galliano2003,galliano2005,engelbracht2005,engelbracht2008,hunt2010}.
In the Milky Way and other metal rich disk galaxies, modeling the dust as being composed of graphites and amorphous silicates accurately reproduces the mid-IR spectral features and far-IR emission \citep{li2001,draine2007}.
Conversely, in metal poor galaxies, modeling grains in this manner
produces unphysical gas to dust ratios \citep[e.g.,][]{bendo2006,galametz2009,meixner2010}, suggesting potential variations in the physical and/or chemical make-up of the dust grains.
\citet{lisenfeld2002}, \citet{meny2007}, and \citet{coup2011} argue for a flattening of the submillimeter slope beyond 500\,$\mu$m, possibly due to non-thermal emission from very small grains with a low emissivity index ($\beta=1$) or changing emissivity properties of larger grains.
An increase in the amount of magnetic material will also change the slope, and hence the derived emissivity, of submillimeter emission \citep{draine2012}.

Complex models of dust emission attempt to take into account a heterogenous grain population and multiple temperature components
\citep[e.g.,][]{dale2002,draine2007, dacunha2008}, but these are difficult to apply when only limited data are accessible. 
The modified blackbody (MBB) is a particularly appealing model due to its simplicity and in most cases gives results identical to more complex modeling or template fitting
\citep[e.g.,][]{magrini2011,galametz2012,casey2012,magdis2012,bianchi2013}. 
It has very few parameters, allowing a determination of temperature when only a few data points are available. 
Dust at far-IR wavelengths primarily radiates thermally as a continuum of blackbodies modulated by the dust grain emissivity.
If far-IR emission is modeled as a single blackbody, in the simplistic assumption of a homogenous optically thin grain population,
the frequency dependence is a power law with a spectral emissivity index, $\beta$, such that
\begin{equation}
S_\nu \propto B_\nu(T_c)\nu^\beta
\end{equation}

MBB modeling is limited in its appeal by the interpretation of the derived parameters. 
When fitting this relationship to data points in order to determine $T_c$ and $\beta$, the
temperature of the blackbody and $\beta$ are anti-correlated and possibly degenerate \citep[e.g.,][]{blain2003,dupac2003,shetty2009,paradis2010,veneziani2010,planck2011,galametz2012,juvela2012}. It has been argued that such an inverse
correlation is non-physical and mostly an effect of detection limits \citep[e.g.,][]{kelly2012}. To avoid uncertainties
introduced by the anti-correlation, published studies commonly advocate holding the emissivity fixed to a standard value in the range $\beta=1-2$ \citep[e.g.,][]{yang2007, xilouris2012,bianchi2013}. 
The intrinsic emissivity is determined by the physical properties of the grains present in individual galaxies. In the Milky Way, it has been measured to be $\beta =1.8$ \citep{planck2011a}. Values of $\beta=2$ are
commonly applied to 
other galaxies as well \citep[e.g.,][]{dunne2001,draine2007}. 

The derived temperature is an average dust temperature. Furthermore, the emissivity derived
contains two pieces of information: first, it is comprised of the intrinsic spectral emissivity index of the dust grains; second, it reflects temperature
mixing within a resolution element, particularly due to dust colder than the peak of the blackbody emission. Multiple dust components, each at a different temperature, can serve to make the effective emissivity shallower than the
intrinsic emissivity of each dust grain population, and even though $\beta$ may change, this does not necessarily
indicate that the composition of the dust is changing.
Therefore, $\beta$ derived from MBB fitting is in reality an effective emissivity.

In the present study, we want to test the robustness of the MBB fitting method and explore how the derived properties of the dust change correlate with the possible heating source of the dust. We build on the work of \citet[hereafter G12]{galametz2012} by modeling the far-IR emission on a resolved scale, using {\it Herschel Space Observatory} images, of a large sample of KINGFISH 
(Key Insights on Nearby Galaxies: a Far Infrared Survey
with Herschel) galaxies. 
Resolved modeling of the far-IR SED is relatively new territory as it requires space-based observations of large galaxies. In recent years, resolved studies with {\it Herschel} have 
mapped dust masses and temperatures, for comparison with gas masses and stellar masses, yielding unprecedented understanding of the ISM in local galaxies 
\citep[e.g.,][]{gordon2010,smith2010,boquien2011a,boquien2011b,aniano2012,bendo2012,foyle2012,mentuch2012,smith2012,draine2014,hughes2014}. 
However, these studies each analyze only one or two objects, although
a large number of pixels (\citet{smith2012} uses 4000 pixels for M~31) allows the closest galaxies to be studied in tremendous detail. G12 improves the number of far-IR resolved studies by probing the distributions of the dust temperatures, emissivities, and dust masses for a sample of 11 KINGFISH galaxies. We now nearly double the sample size of G12 and focus on relating
the distribution of dust temperatures and emissivities, derived from MBB fitting, to the possible heating sources.

\section{Sample Selection}
The KINGFISH sample \citep{kennicutt2012} was selected to include a wide range of luminosities, morphologies, and metallicities in local galaxies. The sample overlaps with 57 of the  galaxies observed as part of the Spitzer Infrared Nearby 
Galaxies Survey \citep[SINGS,][]{kennicutt2003b}, as well as incorporating NGC~2146, NGC~3077, NGC~5457 (M 101), and IC 342 for a total of 61 galaxies. The luminosity range spans 4 orders of magnitude, but all galaxies have $L_{\rm IR} < 10^{11}\,L_\odot$ (with the exception of NGC~2146 which has $L_{\rm IR} \sim 10^{11}\,L_\odot$).
While some of the galaxies display a nucleus with LINER or Seyfert properties, no galaxy's global SED is
dominated by an AGN.

\begin{deluxetable*}{l l c c c c c c c}
\tablecolumns{8}
\tablecaption{Properties of the Sample\,\tablenotemark{a}\label{tbl:properties}}
\tablehead{\colhead{ID} & \colhead{Morph.} & \colhead{D} & \colhead{$R_{25}$\tablenotemark{b}} & \colhead{Size\tablenotemark{c}} & \colhead{Pix Size\tablenotemark{d}} & \colhead{log $L_{\rm TIR}$} & \colhead{SFR$_{\rm H\alpha+24}$} 
 & \colhead{log $M_{\ast}$}  \\
                    \colhead{} & \colhead{} & \colhead{(Mpc)} & \colhead{(arcmin)} & \colhead{(kpc$^2$)} &\colhead{(kpc$^2$)} & \colhead{($L_{\odot}$)} & \colhead{($M_\odot\,{\rm yr}^{-1}$)} & \colhead{($M_\odot$)}}
\startdata
NGC~0628 & SAc	&   7.2  & 5.24	& \, 480	&  0.26	&  9.90	&  0.68	&  9.56	\\
NGC~0925 & SABd	&   9.12	& 5.24	&  \, 800	&  0.70	&  9.66	&  0.54	&  9.49	\\
NGC~1097 & SBb	&  14.2 	& 4.67	&  1500	&  1.4 \,	&  10.7	&  4.2	&  10.5	\\
NGC~1512 & SBab	&  11.6 	& 4.46	& \, 920	&  1.0 \,	&  9.58	&  0.36	&  9.92	\\
NGC~3184 & SABcd	&  11.7	& 3.71	& \, 630	&  0.68	&  10.0	&  0.66	&  9.50	\\
NGC~3351 & SBb	&  9.33	& 3.71	& \, 410	&  0.60	&  9.91	&  0.58	&  10.2	\\
NGC~3521 & SABbc	&  11.2	& 5.48	&  1400	&  1.3 \,	&  10.5	&  2.0	&  10.7	\\
NGC~3621 & SAd	&  6.55	& 6.15	& \, 570	&  0.35	&  9.90	&  0.51	&  9.38	\\
NGC~3938 & SAc	&  17.9	& 2.69	& \, 790	&  1.6 \,	&  10.3	&  1.8	&  9.46	\\
NGC~4254 & SAc	&  14.4	& 2.69	& \, 510	&  1.1 \,	&  10.6	&  3.9	&  9.56	\\
NGC~4321 & SABbc	&  14.3	& 3.71	& \, 950	&  1.1 \,	&  10.5	&  2.6	&  10.3	\\
NGC~4559 & SABcd	&  6.98	& 5.36	& \, 520	&  0.60	&  9.52	&  0.37	&  8.76	\\
NGC~4569 & SABab	&  9.86	& 4.78	& \, 810	&  1.1 \,	&  9.72	&  0.29	&  10.0	\\
NGC~4579 & SABb	&  16.4	& 2.94	& \, 800	&  1.6 \,	&  10.1	&  1.1	&  10.0	\\
NGC~4725 & SABab	&  11.9	& 5.36	&  1400	&  0.94	&  9.94	&  0.44	&  10.5	\\
NGC~4736 & SAab	&  4.66	& 5.61	& \, 230	&  0.12	&  9.76	&  0.38	&  10.3	\\
NGC~5055 & SAbc	&  7.94	& 6.30	& \, 870	&  0.52	&  10.3	&  1.0	&  10.6	\\
NGC~5457 & SABcd	&   6.7   	& 14.4 	&  3000	&  0.22	&  10.4	&  2.3	&  9.98	\\ 
NGC~7331 & SAb	&  14.5  	& 5.24	&  2200	&  3.1 \,	&  10.7	&  2.7	&  10.6	\\ 
NGC~7793 & SAd	&  3.91 	& 4.67	& \, 110	&  0.11	&  9.36	&  0.26	&  9.00	\\
NGC~1291\tablenotemark{e} & SBa	&  10.4 & 4.89	& \, 880 	& 0.60	&  9.54	&  0.35	&  10.8	\\
IC 342\tablenotemark{e}	  & SABcd &  3.28 & 10.7 	& \, 400	& \, 0.051 &  10.1	& 1.87	&  9.95	\\
NGC~3627\tablenotemark{e} & SABb & 9.38  & 4.56 & \, 670 & 0.96 & 10.4  & 1.70  & 10.5  \\
NGC~4826\tablenotemark{e} & SAab	&  5.27	& 5.00 & \, 250 & 0.25 &  9.62	&  0.26	&  9.94	\\
NGC~5474\tablenotemark{e} & SAcd  & 6.8 & 2.39 & \, \, 91 & 0.24  & 8.79 &   0.091  & 8.70 \\
NGC~6946\tablenotemark{e} & SABcd	&  6.8 & 5.74 & \, 520 & 0.25	&  10.9	&  7.12	&  9.96
\enddata
\tablenotetext{a}{All properties are taken or calculated from \citet{kennicutt2012} unless otherwise indicated.}
\tablenotetext{b}{Radius of the major axis at the $\mu_B=25$\,mag\,arcsec$^{-2}$ isophote \citep{deVac1991}.}
\tablenotetext{c}{The physical extent of the galaxy. Calculated from the angular size listed in \citet{kennicutt2012}.}
\tablenotetext{d}{The physical extent of one $14^{\prime\prime}$ pixel.}
\tablenotetext{e}{These galaxies were rejected from the analysis. See Section \ref{sec:fit} for a discussion.}
\end{deluxetable*}

Data observations and reduction are discussed in detail in \citet{engelbracht2010, sandstrom2010}; and \citet{kennicutt2012}. In the present study, 
we use images from the PACS and SPIRE instruments on the {\it Herschel Space Observatory} spanning a wavelength range of $70-500\,\mu$m and MIPS 24\,$\mu$m images from 
the {\it Spitzer Space Telescope} \citep{kennicutt2003b}. The {\it Herschel} observations were intended to probe the cold emission beyond the optical disk,  therefore all galaxy maps are at least 1.5 times the diameter of the optical 
disk.
 
The raw PACS and SPIRE images were processed from level 0 to 1 with {\it Herschel} Interactive Processing Environment (HIPE) v8. The PACS and SPIRE maps
were then created using the IDL package {\it Scanamorphos} v17 \citep{roussel2011}. 
{\it Scanamorphos} is
preferred to HIPE for its ability to better preserve low level flux, reduce striping in areas of high background, and correct brightness drifts caused by low level noise.
The PACS images used for this analysis correspond to the v6 reponsivity calibration in the HIPE context. In order to update to the v7 calibration, we multiply the 70\,$\mu$m, 100\,$\mu$m, and 160\,$\mu$m images by
1.0, 1.0152, and 1.0288, respectively. In addition, the external PACS mappers have released a new calibration for the 160\,$\mu$m images (in HIPE v12), lowering the 160\,$\mu$m flux by $\sim7\%$. Therefore, we also scale the 160\,$\mu$m image
by 0.925 to take this latest calibration into account.

The SPIRE maps are converted from units of Jy/beam to MJy/sr by dividing by beam sizes
of 469.1, 827.2, and 1779.6 arcsec$^2$ for the 250, 350, and 500$\,\mu$m maps, respectively \citep{griffin2013}. The data are then projected onto a grid where one pixel corresponds to 1/4th the beam size of the instrument.
We also conform to the latest HIPE v11 calibration by multiplying the 250, 350, and 500\,$\mu$m images by 1.0321, 1.0324, and 1.0181, respectively.

\begin{deluxetable}{l | cc|c  }
\tablecolumns{4}
\tablecaption{Derived Properties from SED fitting \label{tbl:median}}
\tablehead{\colhead{} & \multicolumn{2}{c|}{$T_c, \beta$ Varied} & \multicolumn{1}{c|}{$\beta$ Const.}  \\
\hline
\multicolumn{1}{c|}{ID} & \colhead{$T_c$ (K)} &\multicolumn{1}{c|}{$\beta$} & \multicolumn{1}{c|}{$T_c$ (K)} }
\startdata
NGC~0628 & 21.4 $\pm$ 1.1 & 1.91 $\pm$ 0.21    & 21.2 $\pm$ 0.92 \\ 
NGC~0925 & 25.6 $\pm$ 1.5 & 1.02 $\pm$ 0.17 \, & 25.6 $\pm$ 1.9 \, \\ 
NGC~1097 & 20.5 $\pm$ 2.1 & 2.02 $\pm$ 0.21    & 20.1 $\pm$ 2.3 \, \\ 
NGC~1512 & 21.1 $\pm$ 1.7 & 1.86 $\pm$ 0.16    & 20.6 $\pm$ 2.1 \, \\ 
NGC~3184 & 20.3 $\pm$ 1.2 & 2.00 $\pm$ 0.21    & 20.1 $\pm$ 0.84 \\ 
NGC~3351 & 20.1 $\pm$ 1.7 & 2.03 $\pm$ 0.15    & 20.0 $\pm$ 1.8 \, \\ 
NGC~3521 & 21.4 $\pm$ 1.1 & 1.85 $\pm$ 0.24    & 20.7 $\pm$ 1.9 \, \\ 
NGC~3621 & 22.8 $\pm$ 1.0 & 1.71 $\pm$ 0.21    & 22.3 $\pm$ 1.8 \, \\ 
NGC~3938 & \, 22.1 $\pm$ 0.71 & 1.84 $\pm$ 0.21 & 21.8 $\pm$ 0.95 \\ 
NGC~4254 & \, 21.4 $\pm$ 0.95 & 2.12 $\pm$ 0.12 & 21.0 $\pm$ 1.1 \, \\ 
NGC~4321 & 20.7 $\pm$ 1.3 & 2.05 $\pm$ 0.11    & 20.4 $\pm$ 1.1 \, \\ 
NGC~4559 & 24.5 $\pm$ 1.2 & 1.26 $\pm$ 0.20    & 24.6 $\pm$ 1.5 \, \\ 
NGC~4569 & 20.2 $\pm$ 1.3 & 2.10 $\pm$ 0.11    & 20.3 $\pm$ 1.0 \, \\ 
NGC~4579 & 19.4 $\pm$ 1.7 & 2.15 $\pm$ 0.18    & 19.2 $\pm$ 1.3 \, \\ 
NGC~4725 & \, 20.2 $\pm$ 0.90 & 1.84 $\pm$ 0.18 & 19.8 $\pm$ 0.86 \\ 
NGC~4736 & 23.1 $\pm$ 2.9 & 1.98 $\pm$ 0.12    & 23.1 $\pm$ 2.6 \, \\ 
NGC~5055 & 19.7 $\pm$ 1.4 & 2.04 $\pm$ 0.21    & 19.1 $\pm$ 1.5 \, \\ 
NGC~5457 & 21.3 $\pm$ 2.3 & 1.79 $\pm$ 0.37    & 21.1 $\pm$ 1.2 \, \\ 
NGC~7331 & 21.1 $\pm$ 1.3 & 1.81 $\pm$ 0.21    & 21.0 $\pm$ 1.8 \, \\ 
NGC~7793 & 24.7 $\pm$ 1.2 & 1.28 $\pm$ 0.20    & 24.6 $\pm$ 1.5 \,\\ 
\enddata
\tablenotetext{}{The values listed are the median and standard deviation for each galaxy.}
\end{deluxetable}

We wish to investigate the temperature and emissivity variations of the dust on spatially resolved scales. For this purpose, we selected all KINGFISH galaxies that are extended
(at least $3.7^\prime\times3.7^\prime$) and are not of the morphological classes E or S0 (as the IR emission of their centrally concentrated star formation is often contaminated by the presence of an AGN); we also remove irregular galaxies to focus our study on the heating of the dust of normal star forming disk galaxies. There are 26 sources that meet these criteria, but we reject six during the fitting process (see Section \ref{sec:fit}), leaving us with 20 galaxies.
The basic properties of these galaxies are listed in Table \ref{tbl:properties}.

\subsection{Background Estimates and Convolution of Images}
The background subtraction is described in detail in \citet{aniano2012}, and we will briefly summarize here. ``Non-background" regions are determined in all cameras (IRAC, MIPS, PACS, and SPIRE)
by those pixels which have a signal-to-noise
ratio (SNR) $>$ 2. All the bands are used to define a mask that contains the emission associated with the galaxy. To remove background emission, the non-masked regions are then fitted by a plane that is finally subtracted from the maps \citep{aniano2012}.

In order to consistently measure photometry for each galaxy, we used images that had been convolved to the resolution of the SPIRE 500\,$\mu$m images. The convolution was done with publicly available kernels from \citet{aniano2011} which transform
the point source functions (PSFs) of individual images to the PSF of the SPIRE instrument at 500\,$\mu$m (FWHM of $38^{\prime\prime}$).
The convolution kernels and methodology are described in detail in \citet{aniano2011, aniano2012}. After 
the convolution to a common PSF, the images for each galaxy are resampled to a standard grid, where each pixel is $\sim 14^{\prime\prime}$. 

 In this study, we derive star formation rates from the 24\,$\mu$m and H$\alpha$ images.
The H$\alpha$ images are ground-based optical images from 
the SINGS and Local Volume Legacy data archive. They have been continuum subtracted, and residual foreground stars have been masked before convolution to the common PSF.

\begin{figure}
\plotone{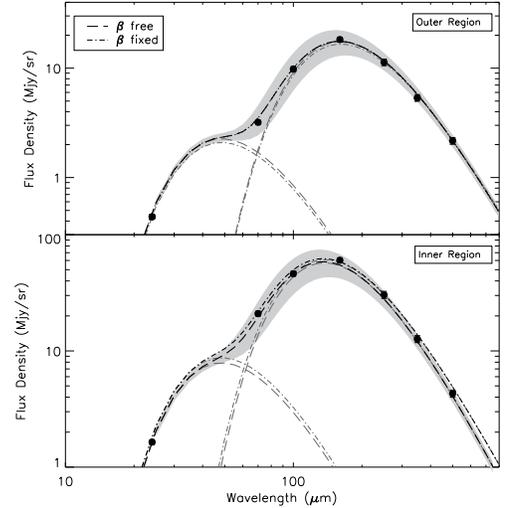}
\caption{We illustrate our 2T MBB fitting method for an outer region and inner region in NGC~0628. The dashed line is the best fit model (and cold and warm dust components) when $T_c$
and $\beta$ vary simultaneously, and the dot-dashed line is the best fit model when $\beta$ is held constant. We show only the errors for the model where $T_c$ and $\beta$ both vary as these are largest. The errors on the individual photometric points are roughly the size of each point. There is a larger discrepancy between the models for the inner region, which we find to generally be the case for the entire sample. \label{fig:example_sed}}
\end{figure}

\section{Modeling the Dust Emission}
\label{sec:fit}
We investigate the spatial variations of the cold dust temperature ($T_c = 15-40$\,K) and the effective emissivity index $\beta$ by modeling the far-IR dust emission on a pixel by pixel basis for each galaxy in our sample. 
We use photometry from $24-500\,\mu$m, in particular MIPS 24\,$\mu$m, PACS 70, 100, 160\,$\mu$m, and 
SPIRE 250, 350, and 500\,$\mu$m images. We fit a two temperature modified blackbody (2T MBB) of the form
\begin{equation}
F_\nu= a_w \times B_\nu(T_w) \times \nu^2 + a_c \times B_\nu(T_c) \times \nu^\beta
\end{equation}
The temperatures of the warm and cold dust components are $T_w$ and $T_c$, the scalings for each component are $a_w$ and $a_c$, and $\beta$ is the emissivity index of the cold dust component.
Only the scalings, $T_c$, and $\beta$ are allowed to be free parameters, due to the limited number of data points being fit.

\begin{figure*}
\centering
\vspace{0.1in}
\includegraphics[width=3.in]{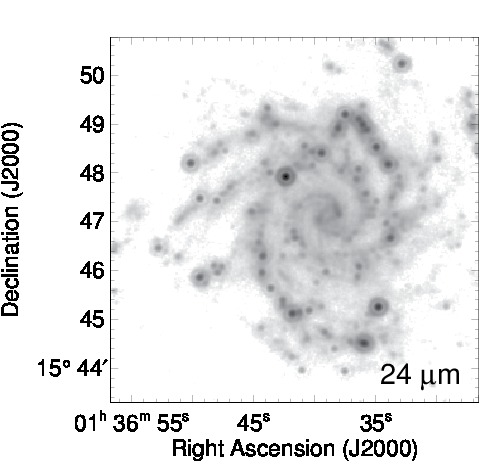}
\includegraphics[width=3.in]{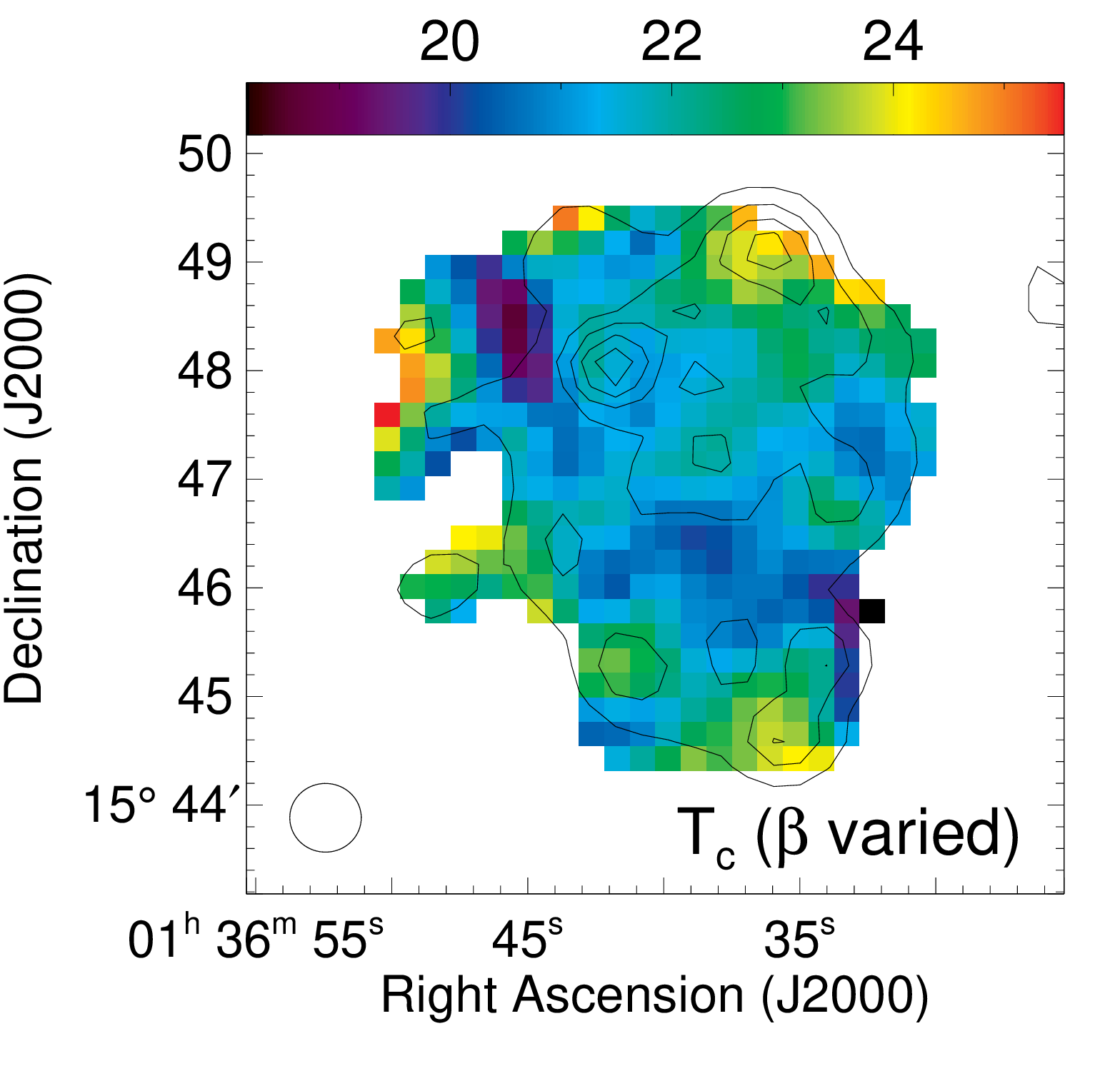}
\includegraphics[width=3.in]{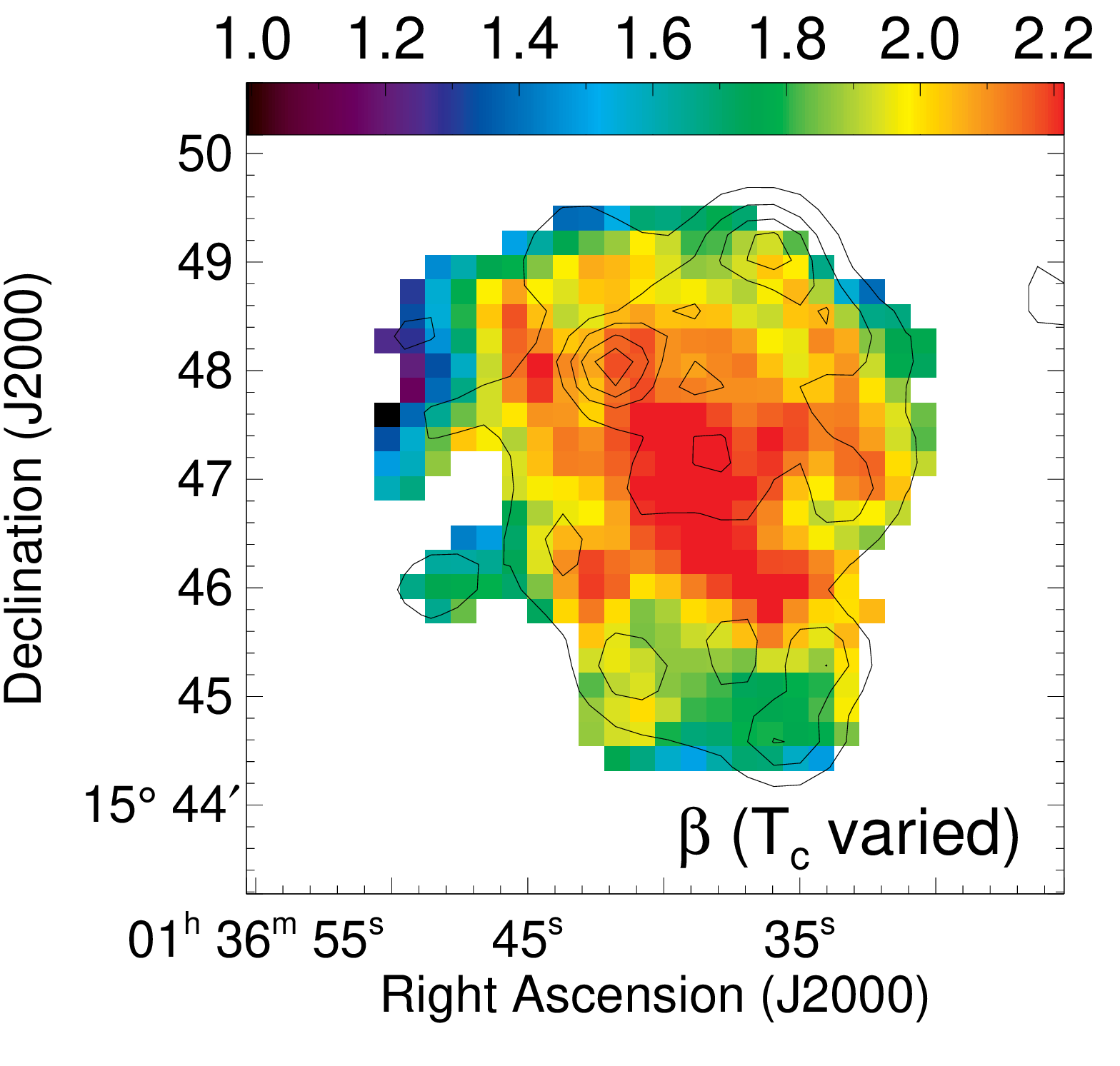}
\includegraphics[width=3.in]{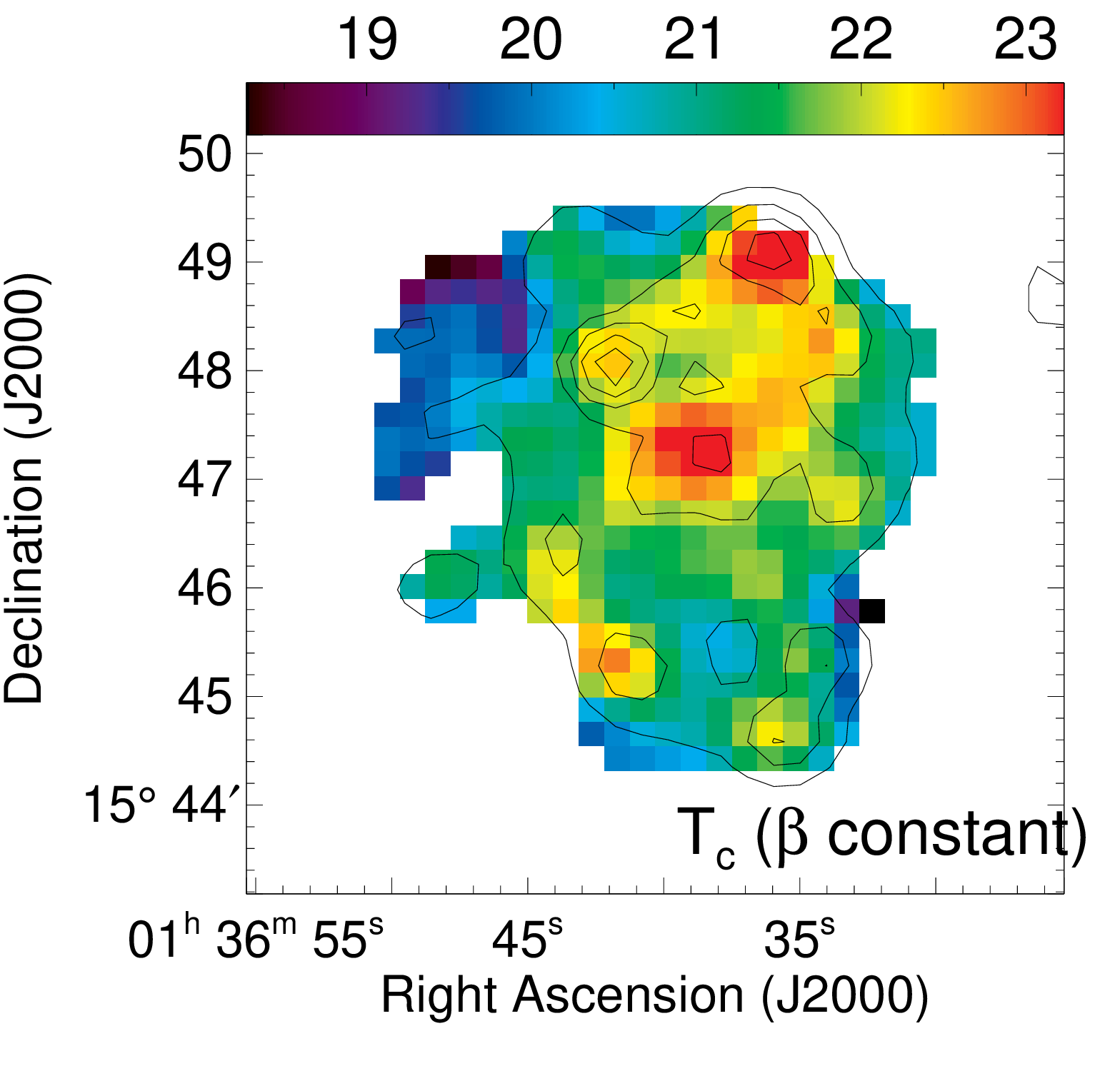}
\caption{Upper left: MIPS 24\,$\mu$m image (MJy\,sr$^{-1}$) of NGC~0628. Upper right: Dust temperature map. $T_c$ is derived when $\beta$ is also a free parameter. The beam size of the SPIRE 500\,$\mu$m image is illustrated in the lower left corner.
Bottom left: Emissivity map for NGC~0628 derived when $T_c$ and $\beta$ are free parameters. Bottom right: Distribution of $\beta$ when $T_c$ is held fixed during the MBB fitting. MIPS 24\,$\mu$m contours (at the resolution of the 500\,$\mu$m image) are overlaid on all $T_c$ and $\beta$ maps. \label{fig:maps}}
\end{figure*}

We add a warm dust component to the model to account for some of the emission at 70\,$\mu$m and 100\,$\mu$m \citep[$\sim15\%$,][]{kirkpatrick2013}. This contamination by warm dust emission usually leads, in the single MBB case, to an overestimation of the dust temperature or a flattening of the emissivity index derived. We experimented with allowing $T_w$ to be a free parameter, but it has a large degree 
of uncertainty due to the sparsity of data points in this range. We fix $T_w$ to 60\,K, which is the median value obtained when we let the warm temperature component vary. 
On average, using a variable warm dust component changes $T_c$ by $-2\,$K and $\beta$ by $+0.25$. Dust in nearby galaxies heated by star formation only has temperatures in the range $15-40$\,K 
\citep[e.g.][]{bendo2003,stevens2005,dale2012}. Warmer dust emission arises from a combination of sources, such as an AGN or an optically thin medium, and a power-law is often a better descriptor of
the emission in this regime than a single MBB \citep{casey2012}. Therefore, we caution that the 60\,K temperature we use should not be interpreted in a strictly physical manner.
We hold the warm dust emissivity fixed to a standard value of 2 which is a good approximation of the opacity of graphite/silicate dust models \citep{li2001}. G12 find that changing $\beta_w$ to 1.5 decreases the cold dust temperatures by less than 1.6\%. By holding the warm dust temperature and emissivity fixed, the warm dust component is in reality serving merely as a normalization for the cold dust component.
We include the 24\,$\mu$m data point in the fit to 
better constrain the warm dust modified blackbody.
 
We have calculated the noise in each image by fitting a Gaussian distribution to the pixels not associated with the galaxy in the background-subtracted images. After calculating SNR for each pixel, we fit the 2T MBB only to the pixels which have a SNR greater than 10 for all wavelengths from 24-500\,$\mu$m since the larger the photometric error bars, the stronger the anti-correlation between $T_c$ and $\beta$ \citep{shetty2009,juvela2012}.

For each spatial location, we randomly resample each data point following a Gaussian distribution within the calibration errors of the instrument (5\% for PACS and 7\% for SPIRE) and refit the model. G12 takes into account correlations between the SPIRE calibrations and only allows the SPIRE photometric data points to vary consistently. We do not do this, but we derive
fitted parameters consistent with G12 nonetheless.
The derived parameters $T_c$ and $\beta$ are the median from a series of 1000 Monte Carlo simulations where we use a $\chi^2$ minimization
to determine the best fit. A Monte Carlo approach is appealing for its simplicity and frequent use 
in the literature,
although there is some indication that this technique will strengthen the degeneracy between $T_c$ and $\beta$ \citep{juvela2013}. We test the goodness of the fits by calculating the residuals at each wavelength. The residuals are largest at 500\,$\mu$m, and at that wavelength, when $T_c$ and $\beta$ are both allowed to vary, we do
not measure any residuals larger than $\sim10\%$ (see discussion in Section \ref{sec:degen}).

We follow two different fitting procedures for each galaxy. First, we allow both $T_c$ and $\beta$ to vary simultaneously, and we will compare these fitted parameters with possible sources of dust heating in 
Sections \ref{sec:heating} and \ref{sec:submm}.
Second, we are interested in how the fitting method used affects the derived cold dust properties. We test what effect fixing a parameter has on 
the resulting temperatures by holding $\beta$ fixed to the median value for each individual galaxy and refitting the 2T MBB allowing only $T_c$ and the 
normalizations to vary. We choose to set $\beta$ to the median value instead of a standard value of 2 because $\beta=2$ does not universally accurately model the slope of the Rayleigh-Jeans tail \citep[e.g.,][]{kirkpatrick2013,galametz2014}. By choosing the median, we are ensuring that any observed trends are not strictly the product of using a poor choice for $\beta$.  Using these techniques, we have 3 maps for each galaxy:
\begin{enumerate}
\item Cold dust temperature map ($T_c$ and $\beta$ are both variable)
\item Cold dust emissivity map ($T_c$ and $\beta$ are both variable)
\item Cold dust temperature map ($\beta$ is held constant)
\end{enumerate}

We illustrate the SED fits in Figure \ref{fig:example_sed} for an inner region and an outer region in NGC~0628. The cold and warm dust components are plotted as the dashed line when both $T_c$ and $\beta$ are derived simultaneously and as the dot-dashed line when $\beta$ is held constant. For clarity, we show only the uncertainty on the fit when $T_c$ and $\beta$ are both derived, 
as this uncertainty is much larger than when $\beta$ is held fixed. There is a larger discrepancy between the two models in the inner region than in the outer region, which we find generally to be the case for the entire sample.

Figure \ref{fig:maps} shows the temperature and emissivity maps for NGC~0628, and the remaining maps are presented in Appendix A. The median $T_c$ and 
$\beta$ values for each galaxy are listed in Table \ref{tbl:median}. It should be noted that the SPIRE beam sizes are larger than the individual pixel size (the beam size diameter is approximately 3 pixels, shown in the upper right panel of Figure \ref{fig:maps}), resulting in $T_c$ and
$\beta$ values that are not independent. We test how this affects the results presented below using NGC~0628, NGC~0925, NGC~4321, NGC~4559, NGC~5055, and NGC~5457. For
these six galaxies, after convolving all images to the 500\,$\mu$m resolution, we then performed photometry with aperture sizes of $42^{\prime\prime}$, slightly larger than the 500\,$\mu$m beam size \citep[for full details, see][]{kirkpatrick2013}. We modeled the photometry with a 2T MBB following the procedure outlined above. We compare the distributions of $T_c$ and $\beta$, derived on a pixel-by-pixel basis and derived after performing aperture photometry, for these six galaxies and find completely consistent results, ensuring that the analysis below is not biased by oversampling the beam.

We remove all pixels with a reduced $\chi^2>>1$. We also examine, on a pixel-by-pixel basis, the distributions of $T_c$ and $\beta$ when allowing both parameters to vary. We remove pixels with bimodal distributions, since 
this makes it difficult to determine the true values of $T_c$ and $\beta$.
We have six galaxies that overlap with the sample in G12. In all cases, our maps are consistent with the maps presented in G12.

We initially modeled 26 galaxies from the larger KINGFISH sample. For clarity, we now describe the galaxies that have been removed from this initial sample. We rejected NGC~1291 because all of the pixels in the 500\,$\mu$m image have an SNR\,$<10$, which is the threshold we 
set 
for determining which pixels in each galaxy would be modeled. NGC~5474 also has a majority of low SNR pixels, leaving only the central kpc to be modeled, which would not allow us to determine how the dust
temperatures and emissivities change in the outskirts of the galaxy.  IC~342 is located close to the Galactic plane, so the Milky Way extinction correction is not uniform across the galaxy. We remove IC 342 to avoid uncertainties introduced by an extinction gradient. 
NGC~3627, NGC~4826, and NGC~6946 have significant regions where the shape far-IR SED is not well described by a 2T MBB with $T_w = 60$\,K. This results in more than 40\% of the spatial regions having bad fits (either a bimodal distribution of $T_c$ and $\beta$ or reduced $\chi^2>>1$).
NGC~3627 and NGC~4826 are
dominated by an AGN in the center \citep{moustakas2010} which could be an underlying cause of the shape of the SED differing from our model.
For NGC~6946, the spatial regions with bad fits also lie in the center of the galaxy, so the powerful central starburst could be responsible. In any case, we wish to apply the same model consistently across all regions of each galaxy, so we opt to remove these three galaxies. In total, we rejected six galaxies from subsequent analysis, leaving us with a final sample of 20 
galaxies.

\begin{figure*}
\centering
\includegraphics[width=6.25in]{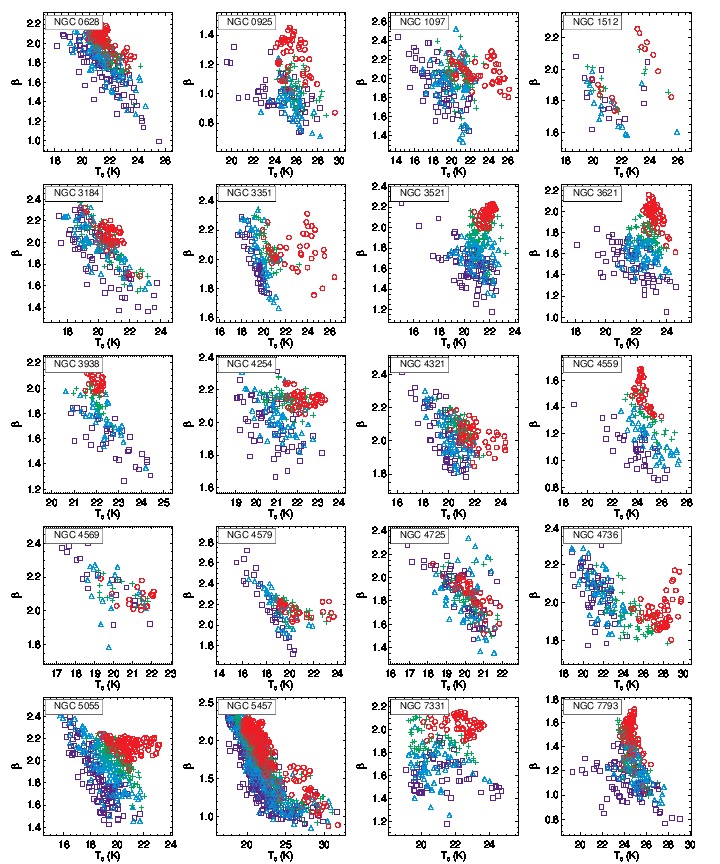}
\caption{The relationship between the derived cold dust temperature and emissivity for each individual galaxy in our sample. The colors correspond to the SNR, with the red circles having the highest SNR and the purple squares having the lowest. In general,
those pixels with a low SNR (purple) display a stronger anti correlation than those pixels with a high SNR (red).
\label{fig:degen}}
\end{figure*}

\begin{figure}
\centering
\includegraphics[width=3in]{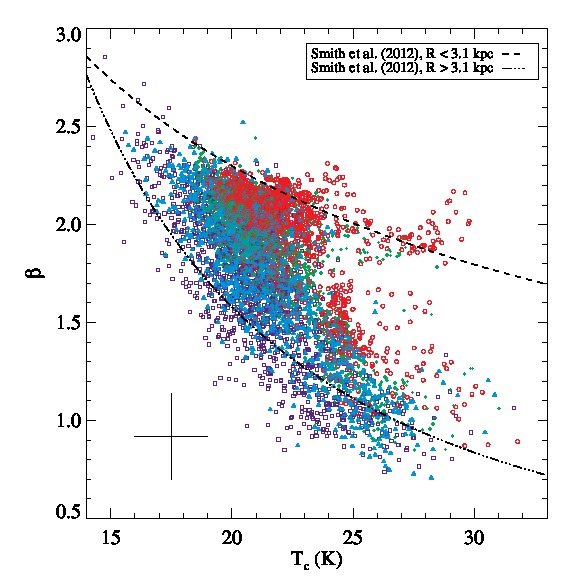}
\caption{The relationship between the derived cold dust temperature and emissivity for all 20 galaxies in our sample. There is an anti-correlation between $T_c$ and $\beta$ that is
slightly stronger for the lower SNR points (purple). The typical uncertainties are shown in the lower left corner.
\label{fig:degen_all}}
\end{figure}

\section{Results}
\subsection{Optimal MBB fitting method}
\label{sec:degen}
The relationship between temperature and emissivity in the individual galaxies is illustrated in Figure \ref{fig:degen}. There is in general an anti-correlation between $T_c$ and $\beta$ that is due, at least in part,
to a degeneracy in MBB fitting. We can use SNRs to test how much of the anti-correlation between $T_c$ and $\beta$ is due to the
fitting method. For each galaxy, we have calculated the 500\,$\mu$m SNR, which is the lowest of all the SNRs, in every spatial location. We then calculate
the median and the lower and upper quartiles for each galaxy. We plot each point according to its SNR quartile, so that red points have the highest SNR and purple points have the lowest. 

We quantify the strength of the anti-correlation between $T_c$ and $\beta$ by calculating the Spearman's rank correlation coefficient for each of the SNR quartiles in each galaxy.
In the lowest quartile, the average correlation coefficient of the 20 galaxies and standard deviation is $\rho=-0.64\pm0.22$, confirming that these $T_c$ and $\beta$ values
are strongly anti-correlated. For the highest SNR data, the average $\rho=-0.28\pm0.39$, which is a weaker correlation, although the standard deviation indicates a large spread in the anti-correlations from galaxy to galaxy. This could indicate that a degeneracy in our sample is strengthened by a low SNR, but since we restrict ourselves to spatial regions with SNR $>$ 10, the observed trends possibly have a physical explanation. The highest SNR regions preferentially trace star forming regions, so the lack of an anti-correlation indicates that the dust properties are more homogenous, while the lower SNR regions are typically tracing the diffuse ISM, where we expect to see more of a 
range of dust temperatures. 

Most galaxies have a centrally located star forming region (we present maps of the star formation regions in Appendix B). The galaxies that have a significant amount of star formation in the spiral arms are NGC~0628, NGC~3184, NGC~4725, and NGC~5457. These are the four galaxies whose distributions of $T_c$ and $\beta$ do not change
significantly from regions of low SNR to high SNR. Conversely, the galaxies with a centrally concentrated star formation region and little star formation in the outskirts, display an offset
between the distributions of $T_c$ and $\beta$ from regions of low SNR to high SNR. The most noticeable offsets occur in NGC~1097, NGC~3351, NGC~4736, and NGC~5055; in our sample, these galaxies exhibit some of the steepest declines in star formation from the center to the outskirts.

In Figure \ref{fig:degen_all}, we plot $T_c$ and $\beta$ for all galaxies together. The anti-correlation between $T_c$ and $\beta$ is now more evident. When
all galaxies are considered together, the Pearson's correlation coefficient for the low SNR points is $\rho=-0.78$ while for the highest SNR points, it is $\rho=-0.59$ (driven primarily by NGC~5457, which has the largest number of pixels) indicating that there is in general a
stronger anti-correlation between $T_c$ and $\beta$ for lower SNR pixels. 
We also measure the two-sided significance for the $\rho$ values. The two-sided significance is a number in the range [0,1] which measures the significance of the 
deviation of $\rho$ from zero, or the null hypothesis. A lower value indicates a higher significance. Both the $\rho$ for the low SNR data and the high SNR data have a two-sided significance consistent with zero, indicative of a real anti-correlation between $T_c$ and $\beta$.

There appears to be two different relationships for $T_c$ and $\beta$. \citet{smith2012} found a similar result for M~31, with $T_c$, $\beta$ values in the inner 3.1 kpc
exhibiting a different trend than in the outskirts. Using an empirical model of $\beta=T^\alpha$ \citep[see also][]{desert2008,paradis2010,planck2011}, the authors find $\alpha=-0.61$ in the inner regions and $\alpha=-1.57$ in the outer regions. We overplot both relations in Figure \ref{fig:degen_all} and find them to be in good agreement with our data set. In general, the points which follow the shallower trend lie primarily within $0.2R_{25}$, where $R_{25}$ is the optical radius, while the points which clearly follow the steeper trend lie beyond 0.4$R_{25}$ (even the highest SNR points), indicating the the two different trends are a product of radius rather than SNR. \citet{smith2012} explains the two different trends with changing properties of $\beta$, which will be discussed further in Section \ref{sec:submm}.

We can examine the 2-D distribution of $T_c$ and $\beta$ as well, to see how the fitting method affects the derived parameters. All $T_c$ and $\beta$ maps are presented in Figure \ref{fig:maps} and Appendix A. To compensate for the degeneracy between $T_c$ and $\beta$, a common technique advocated
in the literature is to hold the emissivity fixed. For our galaxies, we find that holding the emissivity constant or letting
it vary produces marginally worse reduced $\chi^2$ values (higher by $\sim$0.5), but the resulting maps can look quite different (Appendix A), with the strongest variations being in the
central regions, similar to what \citet{foyle2012} found for M~83.
When both parameters are fitted simultaneously, the cold dust temperature maps for 10 galaxies (NGC~1097, NGC~1512, NGC~3351, NGC~4254, NGC~4321, NGC~4569, NGC~4579, NGC~4736, NGC~5055, and NGC~7331)
exhibit a strong radial dependence.
When $\beta$ is held fixed, $T_c$ {\it always} shows a radial dependence. We test if this is a product of the fitting procedure by holding $T_c$ fixed to its median value in each galaxy and allowing $\beta$ to be the free parameter. The resulting maps are indistinguishable from the $T_c$ ($\beta$ constant) maps presented in the Appendix, demonstrating that any radial  temperature variations derived with a fixed 
$\beta$ are not unique, and are induced by the fitting procedure. 

When we hold $\beta$ fixed, the resulting $T_c$ maps illustrate
that the far-IR SED shape
is changing radially, as demonstrated in G12. 
However, we find that it is not necessarily the peak of the SED (quantified by $T_c$) that is decreasing radially in every galaxy. Only 50\% of our sample have similar radial distributions of $T_c$ when $\beta$ is fit as a free parameter and when it is held fixed. In the remaining 50\%,
$\beta$ shows a stronger radial dependence. G12 find that the radial dependence of $T_c$ when $\beta$ is a free parameter corresponds to the presence of a bar, but with our larger sample size, we find that 
this is not the case, and there is no obvious morphology dependence to the distributions of $T_c$ and $\beta$.
With MBB fitting, two parameters are used to quantify the shape of the SED. Because $T_c$ and $\beta$ each show radial trends in galaxies in our sample, we conclude that letting both parameters vary
when fitting an MBB better describes the far-IR dust emission. Holding one parameter fixed forces the other to compensate for the shape of the far-IR SED, inducing a radial trend. G12 recommend holding 
$\beta$ fixed when deriving $T_c$, mainly because they did not see a correlation between the dust temperatures derived when $\beta$ is a free parameter and possible heating sources of the dust for several galaxies. We calculate the heating of the dust in a slightly different manner and discuss possible reasons
for the radially varying far-IR dust emission in Sections \ref{sec:heating} and \ref{sec:submm}.

We can quantify how well the MBB model is fitting the SED by examining any residual emission. The fractional residuals (($S_{500}-{\rm MBB}_{500})/S_{500}$) are largest at 500\,$\mu$m.
When $\beta$ is a free parameter, for all galaxies except NGC~3621 and NGC~4569, the residuals as a function of radius are consistent with 0. 
When $\beta$ is fixed, there is a slight 
anti-correlation between the residuals and radius for 10 galaxies, while a few galaxies also show a positive correlation between residuals and radius, notably NGC~5457. 
The trend in residuals when $\beta$ is constant tells us that the model is not properly tracing the shape of the SED, as it is when both $T_c$ and $\beta$ are variable. Using the nearby galaxy M~31, \citet{smith2012} and \citet{draine2014} also find that on a resolved scale, a variable dust emissivity index is required to properly fit all of the galaxy regions.
Strictly speaking, $\beta$
merely quantifies the slope of the Rayleigh-Jeans tail. As mentioned above, it is an effective emissivity, and so it contains two components: (1) the intrinsic physical emissivity of the dust and (2) the range of temperatures within a
resolution element. Letting $\beta$ vary will
account for the physical properties of the dust grains changing with radius and the amount of temperature mixing varying throughout the galaxy. In the analysis that follows, we consider only the $T_c$ and 
$\beta$ values derived when both parameters are allowed to vary simultaneously.

\begin{figure*}
\centering
\includegraphics[width=6.25in]{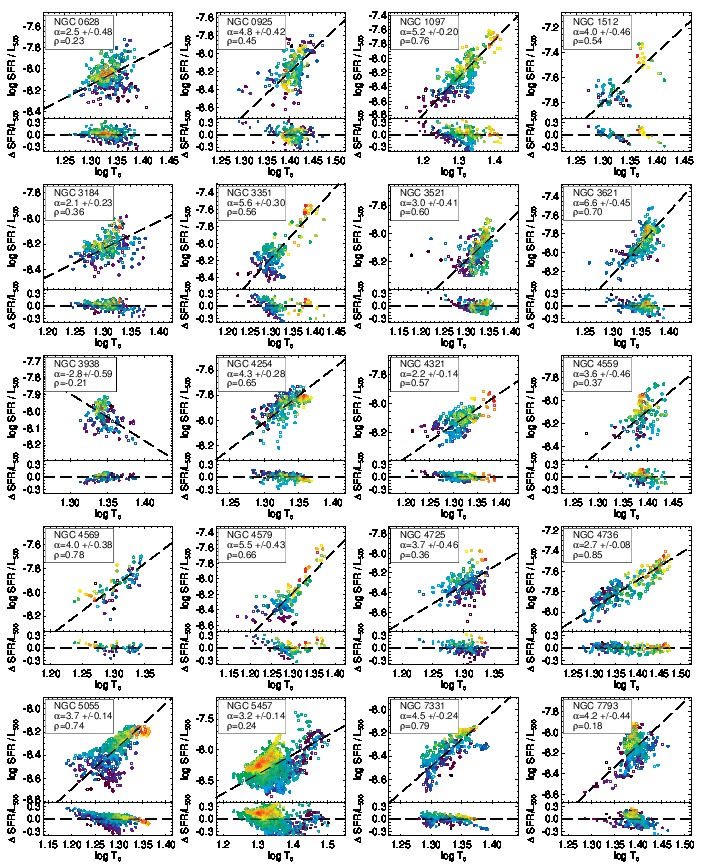}
\caption{The relationship between SFR/$L_{500}$ and $T_c$. There is a positive correlation between the effective SFR and $T_c$ indicating that UV photons from SFR regions are 
responsible for heating the bulk of the diffuse dust. Points are shaded according to normalized radius ($r/R_{25}$) where $r$ has been corrected for the inclination of the galaxy. The radial color scale is illustrated
in Figure \ref{fig:sfrl500_t_all}. \label{fig:sfr_mass_t}}
\end{figure*}

\subsection{Heating Sources of the Cold Dust}
\label{sec:heating}
We now wish to examine the underlying physical relationships between the dust temperature, $T_c$, and the heating of the dust in more detail. The dust is heated primarily through two sources: the young and
the old stellar populations.

\subsubsection{Newly Formed Stars}
 We can trace the heating by the young stars by examining the star formation rate (SFR). We 
calculate the SFR using the H$\alpha$ and 24\,$\mu$m fluxes \citep{calzetti2007}:
\begin{equation}
\left(\frac{\rm SFR}{M_\odot\,{\rm yr}^{-1}}\right) = 5.45 \times 10^{-35} \left(\frac{(L({\rm H}\alpha) + 0.031 \nu L_\nu(24 \mu{\rm m})}{\rm Watts}\right)
\end{equation}
We show maps of the derived SFRs for each galaxy in Appendix B. Because the SFRs are calculated using 24\,$\mu$m emission, these maps indicate regions where the dust is re-radiating light primarily from young photons, although some diffuse emission might also be present. As might be expected, we see a strong similarity between the heating maps and the $T_c$ maps, particularly for NGC~1512, NGC~3351, NGC~4321, NGC~4579, NGC~4736, and NGC~5055, similar to the correlation between resolved 24\,$\mu$m emission and the cold
dust temperature found for NGC 891 \citep{hughes2014}.
However, the differences between the distribution of SFRs and $T_c$ values indicate that for many of our galaxies, SFRs alone cannot fully account for the temperature distributions of the dust.

A more physical comparison can be made between $T_c$ and the effective SFR, that is, the SFR normalized by the dust mass surface density. The effective SFR is the ratio of available ionizing photons to the amount of diffuse dust 
surrounding star forming regions, and it
will give a better sense of how effective the SFR is at heating the dust. Accurately calculating the dust mass surface density requires knowing the absorption opacity, $\kappa$, which depends on $\beta$. Since
$\beta$ varies in different spatial regions of each galaxy, we would introduce a systematic uncertainty by using a constant $\kappa$ \citep{bianchi2013}. We opt to avoid this uncertainty by using $L_{500}$,
the luminosity at 500\,$\mu$m, as a proxy for the dust mass surface density. Now, the 500\,$\mu$m data point is included in the 2T MBB fitting to determine $T_c$ and $\beta$, possibly introducing a degeneracy in any derived relationship between SFR/$L_{500}$ and $T_c$. We calculate the correlation between $T_c$ and $L_{500}$ for all galaxies and find it to be $\rho=-0.2$, which is a weak anti-correlation. Individually, for 17 galaxies, $-0.1 < \rho<0.1$. The remaining three galaxies, NGC~1512, NGC~4569, and NGC~5055, have $\rho\sim0.4$. Given the weak correlations between $L_{500}$ and
$T_c$, we feel confident that using this parameter as a proxy for dust mass surface density is not inducing any false trends between SFR/$L_{500}$ and $T_c$.

Figure \ref{fig:sfr_mass_t} shows a clear trend between $T_c$ and SFR/$L_{500}$. We plot all points according to radius, corrected for the inclination of the galaxy, normalized by the optical radius. We fit a line to the data of the form $\log ({\rm SFR} / L_{\rm 500})\propto\alpha\log T_c$, and we list the $\alpha$ values for each galaxy
in the figure. 
To determine the slope of the line, we use the ordinary least squares (OLS) bisector method detailed in \citet{isobe1990}. In this method, two least squares lines are calculated, one in which $T_c$ is the independent variable and one in which SFR/$L_{500}$ is the independent variable. The resulting relationship is the line bisecting the two least squares lines. This method is more appropriate for our data points than a simple linear regression fit because the choice of independent and dependent variables is not distinct, and because intrinsic scatter rather than measurement errors dominate the uncertainty in determining
the slope of the underlying relationship.

After normalizing by $L_{500}$, we do not observe a systematic trend with radius for all galaxies. To ensure that the observed trend between SFR/$L_{500}$ and $T_c$ is not driven by the radial dependence of each parameter, we also calculate the correlation coefficients between radius and SFR/$L_{500}$ for each galaxy. We find that the galaxies that have the highest correlation between SFR/$L_{500}$ and $T_c$ do not have corresponding high correlations between SFR/$L_{500}$ and radius.
Since $T_c$ plays no role
in the calculation of SFR, we feel confident that the trend we see between SFR/$L_{500}$ and $T_c$ can be interpreted physically. The ratio of UV photons to dust mass is playing a key role
in determining the average dust temperature in a given region.
We further quantify the strength of the correlations between SFR/$L_{500}$ and $T_c$ with a Spearman's rank-order correlation test. We list the Spearman $\rho$ value in each panel in Figure \ref{fig:sfr_mass_t}. Between
SFR/$L_{500}$ and $T_c$, most galaxies have $\rho > 0.5$, and the average (and standard deviation) for all galaxies is $\rho=0.51\pm0.26$. 

For our galaxies, emission at 24\,$\mu$m is the main contributor to the SFR,
and $L_{24}$ is used as a proxy for $L_{\rm IR}$ in calculating the SFR.  $L_{\rm IR}$ is related to temperature as $L\propto T^{4+\beta}$. 
Theoretically, we expect a relationship close to SFR/$L_{500} \propto T^{3+\beta}$. $500\,\mu$m is not a long enough wavelength for the Rayleigh Jeans approximation of $L_{\rm 500}\propto T$ to strictly hold, so the actual dependence between SFR/$L_{500}$ and T should be slightly less than $\alpha=3+\beta$ (analytically, we calculate it to be $\alpha\approx 2.7+\beta$ at 500\,$\mu$m). 
Now, for our galaxies, $\beta$ varies from pixel to pixel, making the expected slope difficult to predict. To compensate, we generate a suite of 2T MBBs spanning a range of temperatures ($15-30$\,K) and 
$\beta$ values ($0.9-2.5$). We calculate $L_{\rm IR}/L_{500}$ for each 2T MBB and use a simple linear regression (appropriate in this case because the temperature is clearly the dependent variable as it was used to generate the blackbodies) to calculate that $\alpha\approx 4$, while the varying $\beta$ values are responsible for the degree of scatter about this line.
Our calculated $\alpha$ values generally span the range $3-6$, in good agreement with our predictions. 

\begin{figure}
\centering
\includegraphics[width=3in]{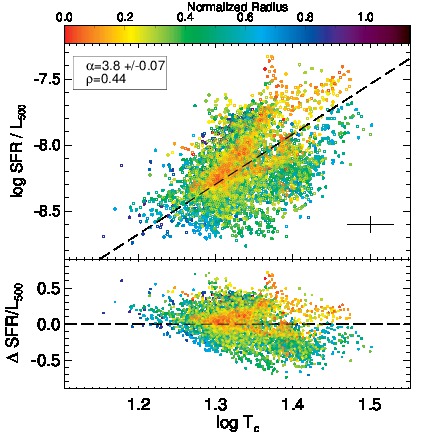}
\caption{The effective SFR, defined as SFR/$L_{500}$ ($M_\odot\,{\rm yr}^{-1}\,L_\odot^{-1}$), as a function of $T_c$ for all the galaxies in our sample. We shade the symbols according to distance from the
center of the galaxy corrected for inclination and normalized by the optical radius. There is a trend with $T_c$, illustrating that the SFR has a strong heating effect on the diffuse ISM. The typical uncertainties are shown in the lower right corner. \label{fig:sfrl500_t_all}}
\end{figure}

Finally, we can examine the resulting relationships when we consider all galaxies in our sample together to look for a systematic trend.
In Figure \ref{fig:sfrl500_t_all}, we plot the effective SFR as a function of $T_c$ for the sample 
as a whole. Again, we fit a line and plot the resulting residuals. We also shade according to the radius normalized by each galaxy's $R_{25}$ value. 
When all galaxies are considered together, $\alpha=3.8$, close to our analytically expected value of $\alpha\approx4$. There is a visible trend between the effective SFR and $T_c$, but the strength of the correlation is only $\rho=0.44$, although the two-sided significance is zero, meaning that the correlation is highly unlikely to be the product of chance. There is also a trend with radius, where pixels lying closer to the center exhibit less scatter around the $\alpha=3.8$ relationship. In general, the mean of the residuals is 0 with a standard deviation of 0.21. However, for the pixels lying within $r/R_{25} \leq0.2$, the standard deviation is 0.16 while for pixels with $r/R_{25} \geq 0.8$, the standard deviation is 0.22.

For our sample of resolved temperatures, SFRs, and luminosities, we see a clear trend between $T_c$ and SFR/$L_{500}$. Globally, a trend between $T_c$ and SFR/$M_{\rm dust}$ is also
observed for a sample of 234 local star forming galaxies
whose global far-IR/submm SEDs have been modeled with the {\sc magphys} code \citep{clemens2013}, illustrating that the trend we see is not merely a byproduct of our 2T MBB fitting method, 
or the fact that we are modeling on a pixel-by-pixel basis. The global relationship between $T_c$ and SFR/$M_{\rm dust}$ for local galaxies indicates that a significant fraction of the cold dust
emission is heated by ongoing star formation, a result which is strengthened by the present analysis where we find that the same correlation holds on smaller scales. For even larger galaxy samples ($>1000$), \citet{dacunha2010} and \citet{smith2012b} find a tight correlation between $M_{\rm dust}$ and SFR, such that galaxies with a larger SFR also have a higher dust mass. A similar result was found on a resolved scale in M~83, where the spatial distribution of the dust mass and SFR correlated strongly, but the dust temperature peaks were offset \citep{foyle2012}. This could explain why we do not observe a strong trend between SFR and $T_c$. As SFR increases, so does the amount of dust present, making it more difficult for the young
stars to heat the dust to higher temperatures. Hence, when we normalize by the amount of dust, we do see a trend with the dust temperature.

If the cold dust is heated solely by the ongoing star formation, then we would expect to see our analytically predicted relation of the
form  SFR/$L_{500}\propto T_c^{3+\beta}$, which is generally observed when all galaxies are considered together. However, because the trend is significantly different than $\alpha=3+\beta$ in some galaxies, and each galaxy has a good degree of scatter, we conclude that ionizing photons leaking out of H{\sc ii} regions are merely 
one source responsible for heating the cold dust.

The temperature derived from MBB fitting tells us the median luminosity-weighted dust temperature. Naturally, we do not expect that all dust within a resolution element is at the same temperature, given the physical size of one of our
pixels. The trend we see between the effective SFR and $T_c$ is indicative of the nature of the ISM; that is, photons leaking out of SF regions are a significant heating source of the diffuse ISM. These photons
determine the average dust temperature, and hence the peak of the SED.

\begin{figure*}
\centering
\includegraphics[width=6.25in]{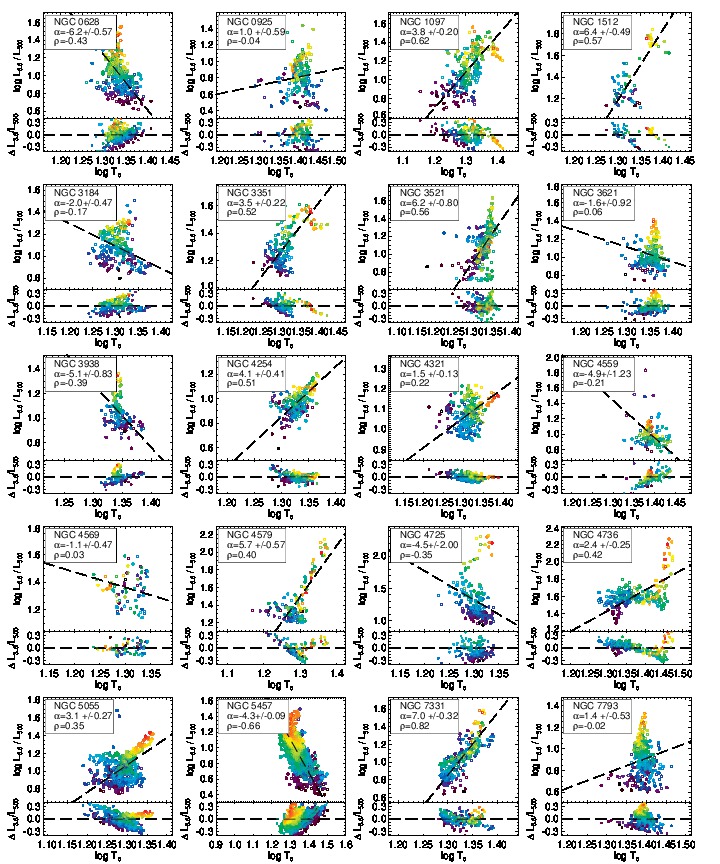}
\caption{The effective heating by the old stellar population, $L_{3.6}/L_{500}$ as a function of $T_c$. We shade the points by normalized radius, with the red points located in the center of the galaxy, and the purple points
in the outskirts. The radial color scale is illustrated
in Figure \ref{fig:l36l500_all}. The correlation between the effective heating by the old stellar population and the dust temperature is generally weak. \label{fig:l36_l500_t}}
\end{figure*}

\begin{figure}
\centering
\includegraphics[width=3in]{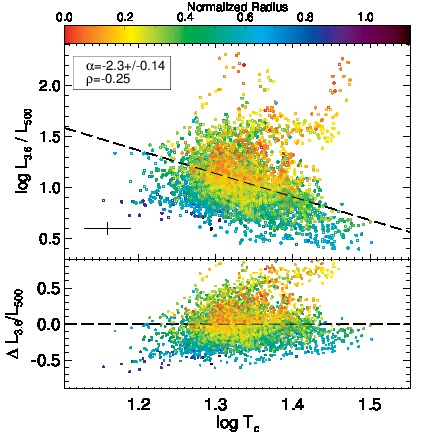}
\caption{The relationship between $L_{3.6}/L_{500}$ and $T_c$ (top panel) and $\beta$ (bottom panel) for all galaxies. We fit a line for all of the galaxies and plot the residuals underneath. We also shade the
points according to radius. These trends are weak and do not appear to be a function of radius. The typical uncertainties are shown in the lower left corner.  \label{fig:l36l500_all}}
\end{figure}

\subsubsection{The Old Stellar Population}
The old stellar population can also contribute to the heating of the diffuse ISM. We explore this avenue using $L_{3.6}$, the luminosity at 3.6\,$\mu$m, as a proxy for the stellar mass. We attribute all of the
luminosity at $3.6\,\mu$m to the old stellar population, though in reality, there can be some contribution from mid-IR emission features here. We expect, however, that any contribution is negligible compared with the 
luminosity due to stars (e.g., \citet{meidt2012} find that intermediate age stars and PAH features account for $\lesssim10\%$ of the integrated light from a galaxy). We show the distribution of $L_{3.6}$ for each galaxy in Appendix B. 

Now, the old stellar population resides in an exponential disk that declines with radius. The dust temperatures are also seen to decline in the outskirts of galaxies. 
The trends between $L_{3.6}$, $T_c$, and radius makes it difficult to determine whether $L_{3.6}$ and $T_c$ are causally related. But, as with SFR, 
we can perhaps get a better hold on the physical underpinnings when we
normalize the photons originating from the old stellar population by the amount of dust there is to heat. The ``effective'' stellar heating is then $L_{3.6}/L_{500}$.

In Figure \ref{fig:l36_l500_t}, we plot $L_{3.6}/L_{500}$ as a function of $T_c$. The effective heating, $L_{3.6}/L_{500}$, has less of a dependence on radius,
though it is still anti-correlated (on average, $\rho=-0.63\pm0.23$). The correlation between $L_{3.6}/L_{500}$ and $T_c$ 
is weaker than the correlation between $L_{3.6}$ and $T_c$, confirming that any correlation between $L_{3.6}$ and $T_c$ is likely driven by a radial distribution of each individual parameter.
When considering the relationship between $L_{3.6}/L_{500}$ and $T_c$, the majority (13) of the galaxies have 
$-0.5<\rho<0.5$, and on average, $\rho=0.14\pm0.42$. However, five galaxies (NGC~0925, NGC~3621, NGC~4559, NGC~4569, and NGC~7793) have a two-sided significance for $\rho$ significantly different from zero, meaning we cannot rule out the null hypothesis that the two parameters are not correlated in these galaxies. In Appendix B, we present
all of the heating maps (SFR, SFR/$L_{500}$, $L_{3.6}$, and $L_{3.6}/L_{500}$) for each galaxy.

We fit a power-law to the data of the form $L_{3.6}/L_{500}\propto T_c^\alpha$ using the OLS bisector method, and we list $\alpha$ in each panel. Both the $\alpha$ and $\rho$ values illustrate that there is
not a universal trend between $L_{3.6}/L_{500}$ and $T_c$. Most galaxies display positive correlations, but a few (e.g., NGC~5457) have anti-correlations.
NGC~0925, NGC~3621, NGC~4559, and NGC~7793 have $\alpha$ values that visually do not seem to correspond to the data. Again, these galaxies also have virtually no correlation between $L_{3.6}/L_{500}$
and $T_c$, so fitting a line to the data is meaningless.

We also look at the trends for the sample as a whole in Figure \ref{fig:l36l500_all}. Again, we shade the points according to normalized radius. The slope of the relationship between 
$L_{3.6}/L_{500}$ and $T_c$ is not strongly influenced by the radial position of the pixels, although there is an offset between the pixels closer to the center and the pixels in the outskirts of the galaxies. In Figure \ref{fig:l36_l500_t}, some galaxies displayed a positive
correlation between the parameters while others displayed an anti-correlation. When considered together, $L_{3.6}/L_{500}$ is weakly (but significantly) anti-correlated with $T_c$ ($\rho=-0.25$), although this trend is driven by NGC~5457.
When we remove NGC~5457, we calculate $\rho=-0.12$ (two-sided significance is still zero). In other words, there does not appear to be a consistent correlation or anti-correlation between $T_c$ and the heating by the old stellar population. 

\begin{figure}
\centering
\includegraphics[width=3in]{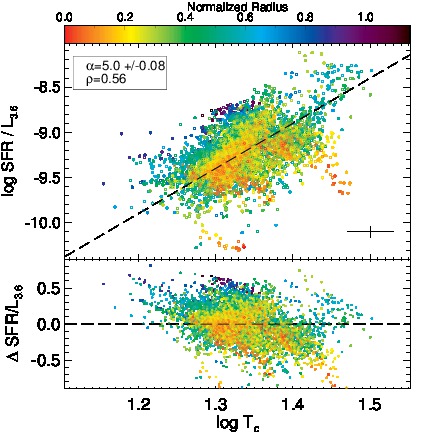}
\caption{SFR/$L_{3.6}$ as a function of $T_c$ (colors correspond to normalized radius). The positive trend with $T_c$ indicates that ionizing photons from newly formed stars are primarily responsible
for determining the average dust temperature (the peak of the SED) in the surrounding diffuse ISM. Typical uncertainties are shown in the lower right corner. \label{fig:sfr_l36_t}}
\end{figure}

Dust emission can be thought of, in a simplistic manner, as arising from two main components: (1) The warm component arises from dust surrounding active regions of star formation or an AGN,
and (2) the cold component is diffuse dust tracing H{\sc i} and heated by the interstellar radiation field (ISRF). 
We can confirm whether young or old stars are the dominant heating source contributing to the peak of the SED by looking at the ratio of SFR$/L_{3.6}$ 
as a function of $T_c$ (Figure \ref{fig:sfr_l36_t}).  SFR/$L_{3.6}$ will tell us the ratio of ionizing photons to less energetic photons from the old stellar population.
Both SFR and $L_{3.6}$ have a radial dependence, but the ratio of the two largely removes this dependence.
SFR/$L_{3.6}$ is positively correlated with $T_c$, as is expected if photons from young stars are more significantly contributing to the heating at the peak of the SED
($\lambda\approx70-160\,\mu$m) than photons from old stars. Nevertheless, there is a degree of scatter (standard deviation of the residuals is 0.24), indicating the complex relationship between the ISRF and ISM. Some of the scatter 
could be attributable to the physical scales being probed. That is, for a correlation to hold between the old stellar population and the diffuse dust, we require that a photon emitted by a star in
a given pixel is absorbed by a dust grain within the same pixel. For some of our galaxies, pixels correspond to physical scales $>0.5\,$kpc, yet we do not see an increase in the amount of scatter for galaxies with a smaller physical pixel scale.

\section{Discussion}
\label{sec:disc}
\subsection{Nature of the $T_c$-$\beta$ Anti-Correlation}
Due to the derivation of $T_c$ and $\beta$ from MBB fitting, a few words need to be said about possible degeneracies.
Because the parameters $T_c$ and $\beta$ are fit simultaneously, it is possible that the anti-correlation between the two parameters is entirely induced by the fitting procedure. 
\citet{yang2007} investigate in detail the anti-correlation between $T_c$ and $\beta$ using a sample of 14 local luminous infrared galaxies (LIRGs). They compare distributions of $T_c$ and $\beta$ obtained
by randomly selecting pairs of $T_c$ and $\beta$ values from appropriate ranges and by calculating $T_c$ and $\beta$ from
a Monte Carlo fitting technique on simulated SEDs to determine the amount of anti-correlation inherent in the
SED fitting procedure. They find that the inherent anti-correlation in the fitting procedure is smaller than the anti-correlation seen in their 14 galaxies,
leading them to conclude that the anti-correlation is physical in nature. More recently, a similar result using Monte Carlo simulations of $T_c$ and $\beta$ was reached for M~31 by \citet{smith2012} and for the Milky Way using a Monte Carlo Markov Chain algorithm by \citet{paradis2010}.

A physical interpretation is also supported by laboratory measurements of amorphous grains. \citet{mennella1998} measure a monotonic increase in the absorption coefficient and decrease of emissivity index of 
amorphous grains as the dust temperature increases from 24 to 295\,K. However, laboratory experiments are currently unable to measure the difference in dust emission in the temperature range $10-20$\,K, which is more of interest here \citep{coup2011}. An inverse $T_c$-$\beta$ relationship can also result from mixtures of different grain populations, particularly if the far-IR/submm
SED contains emission from small grains or ice mantels. Small grains are predicted to have $\beta\sim1$ \citep{seki1980},
while ice mantles have been predicted to have emissivity values as high as $\beta\sim3.5$ \citep{aannestad1975}. Low dust temperatures encourage the formation of ice mantles, leading to higher emissivity
values, while high dust temperatures in very small grains inhibit the formation of ice mantles, leading to lower emissivity values. Therefore, on large scales, if the far-IR emission contains a large
contribution of small dust grains, observers will measure higher $T_c$ values and lower $\beta$ values, while the inverse will be measured if ice mantles dominate the emission.

On smaller scales, the temperature of dust grains is seen to decrease from the diffuse to the dense interstellar medium in the Milky Way, while the emissivity index increases
\citep{schnee2010,cambresy2001,juvela2011,planck2011}. The grain composition also changes; a reduction of mid-IR emission towards dense molecular clouds indicates that small grains are less abundant
\citep{stepnik2003, flagey2009}. Outside the Milky Way, $\beta$ is measured to be lower in the SMC and LMC and shown not to be the result of lower dust temperatures, but instead due to the physical properties of the grain populations\citep{planck2011c}. 
Uncertainties and low SNRs in far-IR photometry can intensify the anti-correlation between $\beta$ and $T_c$ \citep{shetty2009, juvela2012}. \citet{ysard2012} argue that
the observed $T_c$-$\beta$ trend in local molecular clouds is too steep to be explained solely with noise, assuming the cloud has a simple geometry and a dust population with standard optical properties. The
authors therefore conclude that the trend is due to intrinsic variations such as those measured in the laboratory or modeled for amorphous grain emission \citep{mennella1998,meny2007}. 
Alternately, the anti-correlation may in fact be due to temperature mixing along the line of sight \citep{shetty2009}. 
However, by studying local dense molecular clouds, \citet{ysard2012} conclude that such radiative transfer effects cannot explain the observed $T_c$-$\beta$ anti-correlation, since the effect of radiative transfer
is to induce an artificial positive correlation. Furthermore, both \citet{paradis2010} and \citet{anderson2010} demonstrated that the $T_c$, $\beta$ anti-correlation still exists in regions where temperature mixing along the line of sight is unlikely to have an effect.

\begin{figure*}[ht!]
\centering
\includegraphics[width=6.25in]{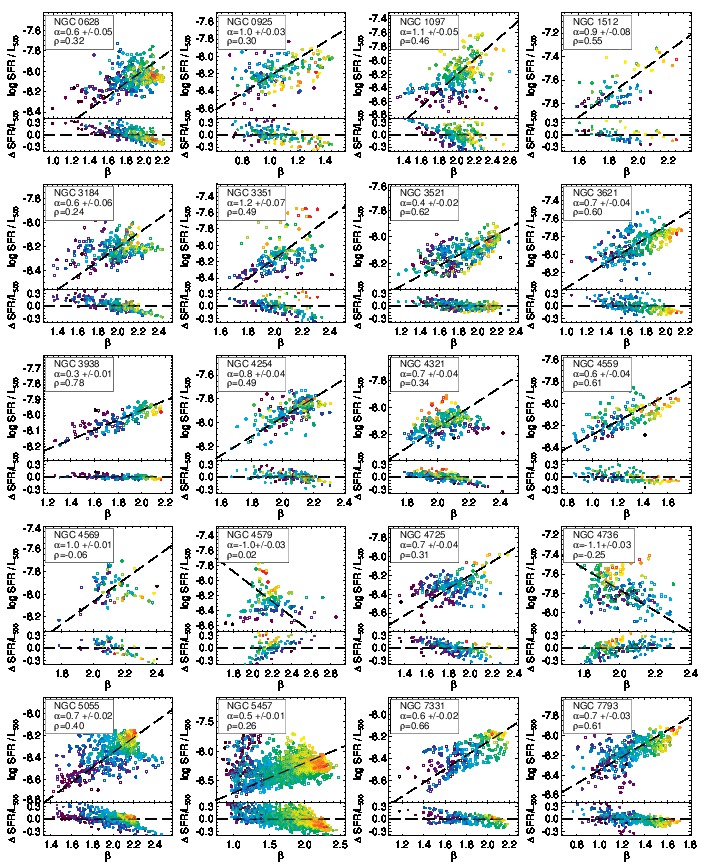}
\caption{The relationship between SFR/$L_{500}$ and $\beta$. There is a positive trend for most galaxies, indicating that the effective SFR
might be contributing to the heating at longer wavelengths. Points are shaded according to normalized radius, and the radial color scale is illustrated
in Figure \ref{fig:sfrl500_both}. \label{fig:sfr_mass_b}}
\end{figure*}

In the $T_c$ and $\beta$ maps for our sample of local star forming galaxies, it is evident that $T_c$ and $\beta$ are not strictly anti-correlated, although this is the general trend, particularly for the low SNR data
(Figure \ref{fig:degen}). In part, we have ameliorated such an effect because of our selection of high SNR 
data points (see Section \ref{sec:fit}).
Two galaxies, NGC~3521 and NGC~3621, exhibit a positive correlation between $T_c$ and $\beta$, while several galaxies (e.g., NGC~4254, NGC~7331) show only a very weak anti-correlation. 
A similar weak anti-correlation has been found at cloud-scale pixel resolution in M 33 \citep{taba2014}. Furthermore, our $\beta$ values display a distinct radial gradient
in 14 of our galaxies, similar to what is observed in other local galaxies \citep[e.g., G12;][]{smith2012,taba2014}. We do not expect that this radial trend is induced by the fitting procedure since it is not observed universally in our sample, and because uncertainties in the data that could be responsible for the $T_c$, $\beta$ fitting degeneracies should be random and not correlate with radius.

The individual trends exhibited for each galaxy have no obvious
correlation with the galaxy's distance or inclination, which would be expected if the strength of the relationship between temperature and emissivity was due strictly to temperature mixing along the line of sight.
We test the effect of temperature mixing by refitting the SEDs of all of our galaxies, this time using a common kpc$^2$ region size, instead of a common pixel size. The use of a standard physical aperture size has 
no effect on any of the trends we have presented above. From this, we conclude that temperature mixing occurs on smaller scales than our resolution elements. Indeed, temperature gradients across molecular 
clouds have been detected locally \citep[e.g.,][]{wang2012}, indicating that temperature mixing occurs even on scales $\lesssim1$\,pc.

Rederiving $T_c$ or $\beta$ while holding the other parameter fixed lends support to the conclusion that the anti-correlation is in part due to the fitting procedure. In this case, the free parameter (either $T_c$ or $\beta$)
exhibits a strongly radial trend. This is due to the fact that the shape of the SED is changing from the center of the galaxy to the outskirts, and when $\beta$ is held fixed, $T_c$ alone must reflect this
change, while when $T_c$ is held fixed, $\beta$ in turn compensates. When the far-IR SED is sparsely sampled, holding $\beta$ fixed is acceptable based on the correlation between the derived temperatures, but caution should be exercised when interpreting any radial trends 
determined with this method. 

\begin{figure}
\centering
\includegraphics[width=3in]{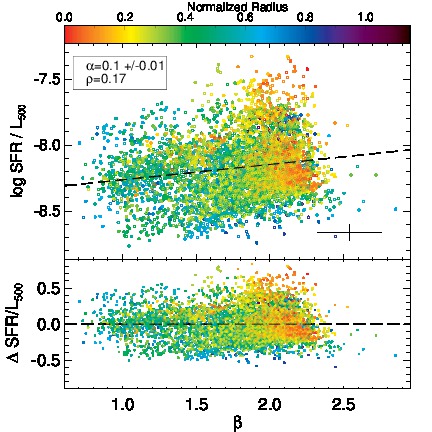}
\caption{The effective SFR, defined as SFR/$L_{500}$, as a function of $\beta$ for all the galaxies in our sample. We shade the symbols according to distance from the
center of the galaxy. There is no obvious trend between the effective SFR and $\beta$. The typical uncertainties are shown in the lower right corner. \label{fig:sfrl500_both}}
\end{figure}

\subsection{The Old Stellar Population and Submillimeter Slope Variations}
\label{sec:submm}
We find an almost universal decrease in the measured $\beta$ as a function of radius for our sources (the notable exceptions are NGC~4254, NGC~4321, NGC~4579, and NGC~4736, where the distribution of 
$\beta$ does not vary greatly with radius). A shallower $\beta$ could reflect a differing mixture of dust grains, but as 
discussed above, the slope of the Rayleigh-Jeans tail is also affected by temperature mixing of dust components within a resolution element. That is, a flattening in the slope of the Rayleigh-Jeans tail
could be explained by the presence of dust components colder than the peak dust temperature.
We can explore the effect of the different dust heating sources on the slope of the Rayleigh-Jeans tail by using the derived effective emissivity to describe the slope.

\begin{figure*}
\centering
\includegraphics[width=6.25in]{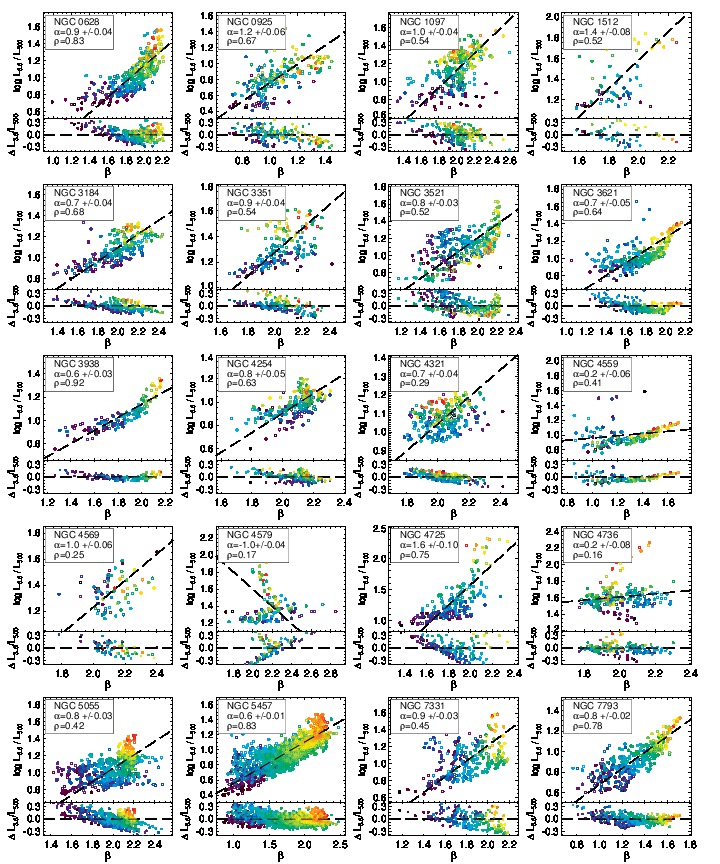}
\caption{The effective heating by the old stellar population, $L_{3.6}/L_{500}$ as a function of $\beta$. We plot according to normalized radius, with the red points located in the center of the galaxy, and the purple points
in the outskirts (the radial color scale is illustrated
in Figure \ref{fig:l36l500_all_b}). There is a strong correlation for most galaxies leading us to conclude that a flatter Rayleigh-Jeans tail is due to inefficient heating, particularly by the old stellar population. \label{fig:l36_l500_b}}
\end{figure*}

\begin{figure}
\centering
\includegraphics[width=3in]{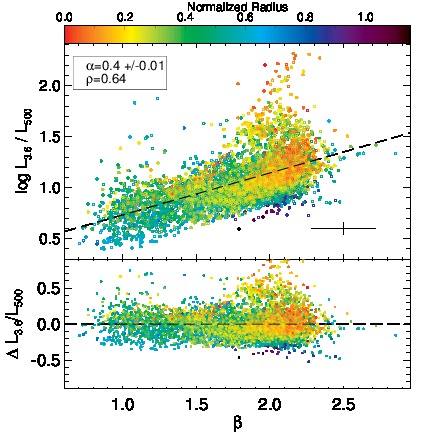}
\caption{The relationship between $L_{3.6}/L_{500}$ and $\beta$ for all galaxies. We fit a line for all of the galaxies and plot the residuals underneath. We also shade the
points according to radius. These trends do not appear to be a function of radius. The trend between $L_{3.6}/L_{500}$ is much stronger with $\beta$ than with $T_c$ illustrating the effect of the old
stellar population on the far-IR/submm slope. The typical uncertainties are shown in the lower right corner. \label{fig:l36l500_all_b}}
\end{figure}

First, we consider the effect of the young stellar population, shown above to be a significant determinant of the average dust temperature.
We illustrate the relationship between $\beta$ and SFR/$L_{500}$ in Figure \ref{fig:sfr_mass_b}. We fit a line to the data of the form $\log {\rm SFR}/L_{500}\propto\alpha\times\beta$ using the OLS bisector method, and we list the $\alpha$ values in each
panel. There is a sub-linear positive trend between $\log {\rm SFR}/L_{500}$ and $\beta$ for most (14) galaxies, although two galaxies have $\alpha<0$.
When we compare $\beta$ and SFR/$L_{500}$, the average
correlation coefficient is $\rho=0.39\pm0.26$, although a few galaxies do have $\rho>0.5$. Despite the small correlation coefficients, for all galaxies except NGC~4569 and NGC~4579, the two-sided significances are consistent with zero. For NGC~4569 an NGC~4579, the two-sided significance does not rule out the null hypothesis that there is
no real correlation. These $\rho$ values are lower than the degree of correlation measured between SFR/$L_{500}$ and $T_c$, indicating that, on average, the effective SFR has a stronger relationship with $T_c$, although
it is still weakly correlated with $\beta$. We show the general trend for all galaxies in Figure \ref{fig:sfrl500_both}. Once all galaxies are considered together, there is a weak, though significant, correlation
($\rho=0.17$, with a two-sided significance of zero), but the scatter around the derived relationship of $\alpha=0.1$ is large (the standard deviation of the residuals is 0.34). In general, there does not appear to be any significant correlation
with radius, although a correlation is present in some individual galaxies (most evident in NGC~5055, NGC~5457, NGC~7331, and NGC~7793). The lack of correlation between SFR and emission in the Rayleigh-Jeans tail is also seen on a resolved scale in M~33 and on a global scale in 51 local galaxies \citep{boselli2010,boquien2011b} where the authors examine a birthrate
parameter $b$ (the SFR normalized by the average SFR over the lifetime of the galaxy) with the far-IR color $S_{350}/S_{500}$. At best, the two parameters are anti correlated, but only weakly,
and the relationship may be affected by the environment of individual galaxies, producing a non-universal trend.

The other prominent source of ISRF photons is the old stellar population. We consider the contribution from the old stellar population in Figure \ref{fig:l36_l500_b}, where we plot $L_{3.6}/L_{500}$ as a function of $\beta$. 
The correlation between $L_{3.6}/L_{500}$ and $\beta$ is strong, with 13 galaxies having $\rho>0.5$. All galaxies are considered together in Figure \ref{fig:l36l500_all_b}, and there is a significant correlation between the parameters, with a Spearman's rank $\rho$ of 0.64 and a two-sided significance of zero, which is stronger than the correlation between $L_{3.6}/L_{500}$ and $T_c$.

$\beta$ is correlated with radius, decreasing towards the outskirts. Again, we test that the correlation between the older stellar population and $\beta$ is not driven solely by radial trends by deriving the correlation coefficients for $L_{3.6}/L_{500}$ and radius for each galaxy. We find that the galaxies that have the highest correlation between $L_{3.6}/L_{500}$ and $\beta$ do not necessarily have the highest correlations between $L_{3.6}/L_{500}$ and radius.

Our correlation between $L_{3.6}/L_{500}$ and $\beta$ is in line with what is found by \citet{boquien2011b} and \citet{bendo2012} for M~33, M~81, M~83, and NGC~2403. In those galaxies, the authors look at the distributions of the
warm dust (traced by the colors $S_{24}/S_{160}$ and $S_{70}/S_{160}$) and the cold dust (traced by $S_{250}/S_{500}$). The authors find significantly different distributions of the two colors, implying that the two dust
components are heated by different sources. In particular, $S_{250}/S_{350}$ shows a tight correlation with either stellar mass (in M~33) or 1.6\,$\mu$m emission, revealing that the old stellar population is driving the emission in this regime. $S_{250}/S_{350}$ increases with increasing stellar mass, meaning that the slope Rayleigh-Jeans tail is steepening; this corresponds to the larger $\beta$ values that we measure as
$L_{3.6}/L_{500}$ increases.

\citet{taba2014} measure a radially decreasing $\beta$ in M~33, and the authors find a correlation between $\beta$ and H$\alpha$ emission, a SFR tracer, leading them to conclude that the physical properties of the dust grains are actually changing due to star formation replenishing the ISM with metals, increasing the amount silicates which have a lower emissivity than carbonaceous grains. We test this conclusion with our larger sample by comparing $\beta$ directly with H$\alpha$. Only seven galaxies have a strong correlation ($\rho > 0.5$), and five galaxies have no significant correlation ($\rho < 0.2$). We find the same results when we compare with SFR$_{{\rm H}\alpha+24\,\mu {\rm m}}$. We compare
this with the results for $L_{3.6}/L_{500}$ and $\beta$, where 12 galaxies have a strong correlation ($\rho > 0.5$), and only two galaxies have no significant correlation ($\rho < 0.2$). 

\citet{smith2012} also argues for a physical interpretation in the variations of $\beta$ measured in M~31. They observe a decrease in $\beta$ in the center of the galaxy and an increase out to 3.1 kpc. The authors conclude that the decrease of $\beta$ in the center maybe due to sublimation caused by an increased ISRF, and the increase of $\beta$ could be caused by the growth of icy mantles.

Metallicity is also a tracer of the physical grain properties, although we do not think that metallicity is solely driving the radial dependence of $\beta$ because the metallicity range in any galaxy is very small ($\log({\rm O/H})+12=8.2-8.8$). In this metallicity range, the Rayleigh-Jeans tail is not typically observed to flatten \citep[e.g.,][]{galametz2011}. The correlation between $\beta$ and H$\alpha$ and SFR$_{{\rm H}\alpha+24\,\mu {\rm m}}$ indicates that the intrinsic grain properties may be driving the radial distribution of $\beta$ in a few galaxies, although the correlation between $\beta$ and the ISRF as quantified by $L_{3.6}$ hints that the variations in $\beta$ are could also be due to inefficient heating in the outskirts of the galaxies allowing for colder dust components beyond the peak of the dust emission.

When colder temperature components are seen beyond the peak of the IR SED, this will be reflected by a shallower emissivity. Low $\beta$ values
correspond to low $L_{3.6}/L_{500}$ values. Higher $\beta$ values, meaning {\it less} temperature mixing, correspond to higher $L_{3.6}/L_{500}$ values. When there are many photons available from the old stellar population to heat the dust, the dust displays a $\beta$ closer
to what might be physically expected from the intrinsic properties of the dust grains alone, meaning there is little dust colder than the peak of the SED. However, when there are few photons available from the older stellar population per unit dust mass, i.e. inefficient heating, the dust has more components at
colder temperatures, leading to a shallower Rayleigh-Jeans slope and a lower derived effective emissivity. The flattening of the Rayleigh-Jeans tail shows a radial dependence in most galaxies in our sample. Recently, \citet{galametz2014} has modeled the resolved submillimeter emission by including 870\,$\mu$m photometry in a sample of local galaxies. When including the longer wavelength data, the authors also see a radial flattening of the Rayleigh-Jeans tail. One possible explanation for a flatter Rayleigh-Jeans tail is a colder dust component, illustrating that heating is the most inefficient in the outskirts of galaxies, which is also supported by the radial decrease of the heating due to young stars and the old stellar population.

\section{Conclusions}
We have explored the resolved temperature and emissivity values of dust emission in a large sample of local star forming galaxies. We model the far-IR dust emission using two temperature modified blackbodies.
We calculate $T_c$ and $\beta$ for the cold dust component by letting both parameters vary simultaneously. We compare the derived parameters with the SFR$_{{\rm H}\alpha + 24\mu{\rm m}}$, $L_{500\,\mu{\rm m}}$, and $L_{3.6\,\mu{\rm m}}$.
We summarize our findings below.
\begin{enumerate}
\item We further the work of G12 by doubling the sample size and performing similar 2T MBB fitting on resolved scales. G12 quantify the dust heating using 24\,$\mu$m emission and 3.6\,$\mu$m emission and in several galaxies, find little correlation between heating and dust temperatures. We improve on the heating measurements by using SFR$_{{\rm H}\alpha +24\mu{\rm m}}$ and normalizing by amount of dust present, quantified by $L_{500}$. We find good agreement between the sources of dust heating and the derived dust temperatures.
\item We advocate letting both parameters vary simultaneously, 
while taking care to remember that the derived parameters are the average dust temperature and
effective emissivity. When one parameter is held constant, the other exhibits a strong radial trend, indicating that the shape of the far-IR SED is changing from the center to the outskirts of a galaxy. In the present analysis, we find that the radial dependence of the dust emission corresponds to the changing radiation field and sources of heating.
\item We find a correlation between the effective SFR (SFR/$L_{500}$) and $T_c$, indicating that the {\it relative} number of photons from young stars has a significant heating effect on the diffuse, cold dust component.
\item We find a correlation between the effective heating by the old stellar population ($L_{3.6}/L_{500}$) and $\beta$, which describes the slope of the Rayleigh-Jeans tail. The Rayleigh-Jeans tail flattens primarily towards the outskirts of galaxies, and this flattening could be due to the presence of colder dust components beyond the peak of the SED. We interpret the observed correlation between $L_{3.6}/L_{500}$ and the derived emissivity to mean that the Rayleigh-Jeans tail flattens due to inefficient heating by the old stellar population.

\item The relationship between effective SFR and $T_c$ shows a degree of scatter for each galaxy which suggests that UV photons are not the only source of heating responsible for determining the peak of the IR SED emission. In turn, the relationship
between $L_{3.6}/L_{500}$ and $\beta$ also shows a degree of scatter. Both the old stellar population and newly formed stars are responsible for heating at all far-IR wavelengths, but the newly formed
stars seem to be driving the peak of the IR SED as suggested by the good correlation between the effective star formation and the temperatures we derive.
\end{enumerate}
\acknowledgements
We thank the anonymous referee for his/her valuable insights which have improved the clarity and quality of this paper.
Herschel is an ESA space observatory with science instruments
provided by European-led Principal Investigator consortia
and with important participation from NASA. IRAF, the
Image Reduction and Analysis Facility, has been developed by
the National Optical Astronomy Observatories and the Space
Telescope Science Institute.
This research has made use of the NASA/IPAC Extragalactic Database (NED) which
is operated by the Jet Propulsion Laboratory, California Institute of Technology, under
contract with the National Aeronautics and Space Administration. FT acknowledges the DFG grant TA 801/1-1.

\end{document}